\documentclass[useAMS,usenatbib,usegraphicx,a4paper]{mn2e}
\usepackage{times}
\usepackage{amssymb}  

\title[Impact of inner rims of protoplanetary discs on Spectral Energy Distributions]{Radiation thermo-chemical models
  of protoplanetary discs. III. Impact of inner rims on Spectral
  Energy Distributions}
\author[W.-F. Thi et al.]{W.-F. Thi$^{1,2}$ \thanks{E-mail: wfdt@roe.ac.uk}, P. Woitke$^{3,1,2,4}$, I. Kamp$^5$\\
  $^1$\thanks{Scottish Universities Physics Alliances}SUPA, Institute for Astronomy, University of Edinburgh, Royal Observatory, Blackford Hill, Edinburgh, EH9 3HJ, UK\\
  $^2$Universit\'e Joseph-Fourier – Grenoble 1/CNRS, Laboratoire d’Astrophysique de Grenoble (LAOG) UMR 5571, BP 53, 38041 Grenoble Cedex 09, France\\
  $^3$UK Astronomy Technology Centre, Royal Observatory, Edinburgh, Blackford Hill, Edinburgh EH9 3HJ, UK\\
  $^4$ School of Physics \& Astronomy, University of St.~Andrews, North Haugh, St.~Andrews KY16 9SS, UK
\\ 
  $^5$Kapteyn Astronomical Institute, Postbus 800, 9700 AV Groningen, The Netherlands\\
}
  
\begin{document}

\date{Accepted 2010. Received 2010}

\pagerange{\pageref{firstpage}--\pageref{lastpage}} \pubyear{2010}

\maketitle
\label{firstpage}
\begin{abstract}
  We study the hydrostatic density structure of the inner disc rim
  around Herbig~Ae stars using the thermo-chemical hydrostatic code
  {\sc ProDiMo}.  We compare the Spectral Energy Distributions (SEDs)
  and images from our hydrostatic disc models to that from prescribed
  density structure discs. The 2D continuum radiative transfer in {\sc
    ProDiMo} includes isotropic scattering. The dust temperature is
  set by the condition of radiative equilibrium. In the
  thermal-decoupled case the gas temperature is governed by the
  balance between various heating and cooling processes. The gas and
  dust interact thermally via photoelectrons, radiatively, and via gas
  accommodation on grain surfaces.  As a result, the gas is much
  hotter than in the thermo-coupled case, where the gas and dust
  temperatures are equal, reaching a few thousands K in the upper disc
  layers and making the inner rim higher. A physically motivated
  density drop at the inner radius (``soft-edge'') results in rounded
  inner rims, which appear ring-like in near-infrared images.  The
  combination of lower gravity pull and hot gas beyond $\sim$~1~AU
  results in a disc atmosphere that reaches a height over radius ratio
  $z/r$ of 0.2 while this ratio is 0.1 only in the thermo-coupled
  case. This puffed-up disc atmosphere intercepts larger amount of
  stellar radiation, which translates into enhanced continuum emission
  in the 3--30~$\mu$m wavelength region from hotter grains at
  $\sim$~500~K. We also consider the effect of disc mass and grain
  size distribution on the SEDs self-consistently feeding those
  quantities back into the gas temperature, chemistry, and hydrostatic
  equilibrium computation.
\end{abstract}  


%
\section{Introduction}\label{rim_intro}

Circumstellar discs result from the conservation of angular momentum
of a collapsing cloud that forms a star at its centre. Discs of
various masses and shapes are observed by direct imaging or
spectroscopically from low mass brown dwarfs (e.g.,
\citealt{Mohanty2004ApJ...609L..33M,Jayawardhana2003AJ....126.1515J,Natta2001A&A...376L..22N})
to massive O stars \citep{Cesaroni2007prpl.conf..197C}. Spectral
Energy Distributions (SED) are dominated by dust grain emission and
are influenced by disc geometry, dust mass, and grain properties.
Herbig~Ae stars (2--4 M$_{\odot}$) are among the most studied
pre-main-sequence stars surrounded by discs because of their isolated
nature and intrinsic brightness
\citep{Natta2007prpl.conf..767N}. Their SEDs can be classified into
two groups according to the ratio between the far-infrared to
near-infrared fluxes \citep{Meeus2001A&A...365..476M}. Group I objects
show much stronger far-infrared to near-infrared fluxes than their
group II counterparts. Early versions of the so-called puffed-up rim
model succeeded to explain the dichotomy
\citep{Dullemond2001ApJ...560..957D,Dominik2003A&A...398..607D}.
Group II objects are discs that absorb most of the stellar flux within
a geometrically thin but optically thick inner rim that faces directly
the star, depriving the outer disc of radiation. As a consequence the
cold outer disc is geometrically flatter and emits only weakly in the
far-infrared. On the contrary, the rim in group I objects does not
block enough stellar radiation and the outer disc flares, emitting
strongly in the far-infrared. 2D dust radiative transfer models that
include dust scattering and hydrostatic disc structure assuming that
the gas temperature is equal to the dust temperature, which is derived
from dust energy balance, underestimate the flux in the near-infrared
around 3~$\mu$m
\citep{Meijer2008A&A...492..451M,Vinkovic2006ApJ...636..348V}. \citet{Vinkovic2006ApJ...636..348V}
invoke the presence of an optically thin halo above the inner disc to
explain the excess 3$\mu$m emission. The problem arises from the
insufficient height of the inner rim: a low rim results in a small
emitting area. \citet{Acke2009A&A...502L..17A} require that their
inner rims have to be 2 to 3 times higher than the hydrostatic
solution that fits the SEDs using a Monte-Carlo radiative transfer
code. Models that incorporate detailed physics of grain evaporation
with grain size distribution change the structure of the inner rim and
help to fit simultaneously the SED and near-infrared interferometric
data (e.g.,
\citealt{Isella2006A&A...451..951I,Tannirkulam2007ApJ...661..374T,Kama2009A&A...506.1199K}). The
structure of the inner rim impacts the overall shape of the SEDs up to
a few tens of microns. Initially discovered in the SED of Herbig~Ae
discs, inner rim emission is also detected in discs around classical
T~Tauri stars \citep{Muzerolle2003ApJ...597L.149M}.  Observationally,
the inner disc geometry is varied. Studies with sufficient
baseline-coverage show that either an extra optical thin component is
required
\citep{Tannirkulam2008ApJ...689..513T,Benisty2010A&A...511A..75B} or
that the rim has a skewed asymmetric geometry
\citep{Kraus2009A&A...508..787K}. A detailed modelling of these
observations is beyond the scope of this paper.

All previous studies of inner disc rims share the assumption that the
gas and dust temperatures are equal to simplify the radiative
transfer. In these studies, the temperatures are derived from the dust
energy balance. This assumption is valid in the inner optically thick
parts of disc midplanes where gas and dust are thermally coupled via
inelastic collisions, a phenomenon called thermal accommodation
\citep{Tielens2005pcim.book.....T}. In all other parts, the gas is
mostly heated by hot photoelectrons from dust grains and Polycyclic
Aromatic Hydrocarbons and attains higher temperatures than the
dust. The inner rim is naturally higher than when the gas and dust
temperature are equal and the overall disc surfaces are more
vertically extended \citep*{Woitke2009A&A...501..383W}. In discs where
the gas and dust are well mixed (i.e. without dust settling), the
change in geometry can affect the SEDs.

In this paper we explore the possibility that inner rims are much
higher when the gas and dust temperatures are computed separately and
self-consistently.  A detailed study of inner and outer
thermo-chemical disc structures together with emission lines in
Herbig~Ae discs is presented in \citet{Kamp2010A&A...510A..18K}. In
this paper we focus on the inner disc structure of well-mixed discs
and the images and spectral signature of inner rims in the infrared.
\begin{figure*}
\centering  
  {\includegraphics[angle=0,width=18cm]{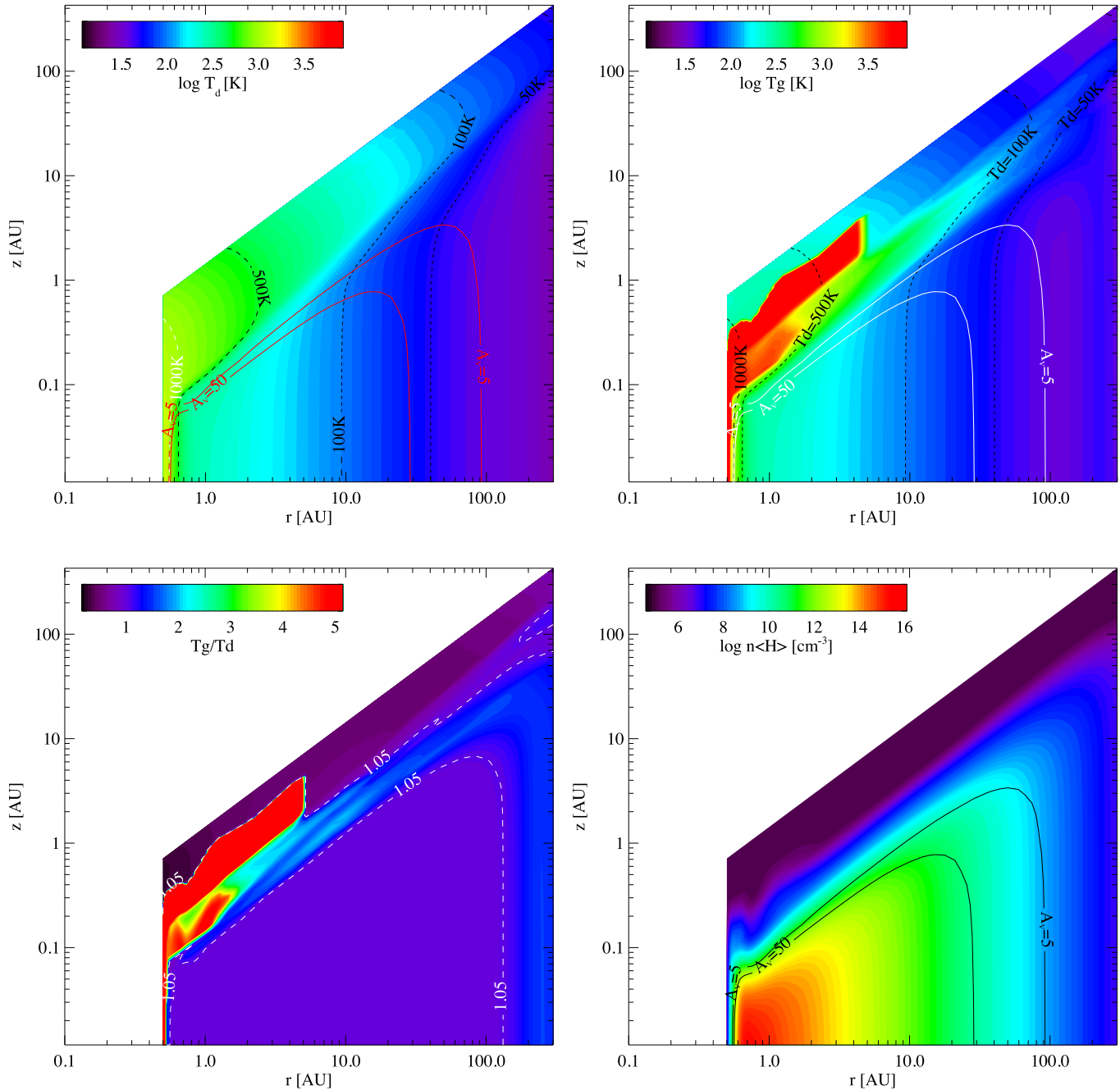}}
  \caption{The upper right and left panels show the dust and gas
    temperature structures respectively for the fiducial model. The
    dust temperature contours are overplotted for $T_{\mathrm{dust}}=$
    50, 100, 500, and 1000~K in both panels. The location of
    $A_{\mathrm{V}}=$~5 and 50 are also shown. The entire outer disc
    at $R>100$~AU is located at $A_{\mathrm{V}}<$~5. The lower-right
    panel show the disc density structure for the entire disc (up to
    $R_{\mathrm{out}}=$~300~AU).  The gas temperature over dust
    temperature ratio for the fiducial model in the lower-left
    panel. The gas can be more than $\sim$~5 times warmer than the
    dust in the upper disc atmosphere but
    $T_{\mathrm{gas}}=T_{\mathrm{dust}}$ at
    $A_{\mathrm{V}}>(1-5)$. The ratio
    $T_{\mathrm{gas}}/T_{\mathrm{dust}}$ is explicitly truncated at 5
    in the figure.}
  \label{fig_fiducial} 
\end{figure*}


This paper is organized as follow. We briefly describe the {\sc
  ProDiMo} code and recent additions to the code such as the SED and
image calculation in Sect.~\ref{prodimo}. We proceed with a
presentation of our fiducial disc model, the disc parameters that are
varied, and the fixed-structure disc model in
Sect.~\ref{model_description}. We then present and discuss the results
of our simulations when the gas and dust temperatures are equal
(thermal-coupled models) and when they are computed self-consistently
(thermal-decoupled models), varying several disc parameters in
Sect.~\ref{results_discussion}. Finally, we conclude about the inner
disc structures and their effects on the SEDs in
Sect.~\ref{conclusion}.

\begin{center}
\begin{table*}
  \caption{Disc parameters. When a parameter has multiple entries, the values in bold correspond to the values of the fiducial model.}\label{tab_DiscParameters}
		\begin{tabular}{lll}
                  \hline
                  stellar mass & $M_*$ &  {\bf 2.2}~M$_\odot$ \\ 
                  stellar luminosity &$L_*$  &  {\bf 32}~L$_\odot$ \\ 
                  effective temperature & $T_{\mathrm {eff}}$ & {\bf 8600}~K\\
                  disc mass & $M_{\mathrm{disc}}$ & {\bf 10}{\boldmath $^{-2}$}, 10$^{-3}$, 10$^{-4}$ M$_\odot$ \\ 
                  disc inner radius & $R_{\mathrm {in}}$  & {\bf 0.5}, 1 , 10~AU\\
                  disc outer radius & $R_{\mathrm {out}}$  & {\bf 300}~AU\\
                  vertical Column density power law index & $\epsilon$ & 1, 1.5, {\bf 2}\\
                  gas to dust mass ratio           &  $\delta$ & {\bf 100}\\
                  dust grain material mass density & $\rho_{\mathrm{dust}}$ & {\bf 2.5} g cm$^{-3}$ \\
                  minimum dust particle size       & $a_{\mathrm{min}}$ & {\bf 0.05} $\mu$m\\
                  maximum dust particle size       & $a_{\mathrm{max}}$ & 10, {\bf 50}, 200 $\mu$m\\
                  dust size distribution power law & $p$               & {\bf 3.5}\\
                  H$_2$ cosmic ray ionization rate       & $\zeta_{\mathrm{CR}}$  & {\bf 1.7 $\times$ 10}{\boldmath $^{-17}$} s$^{-1}$\\
                  ISM UV field w.r.t. Draine field  & $\chi$          & {\bf 0.1}\\
                  abundance of PAHs relative to ISM & $f_{\rm PAH}$      & {\bf 0.01}, 0.1\\
                  $\alpha$ viscosity parameter      & $\alpha$           & {\bf 0.0}\\
                  \hline
\end{tabular}
\ \\ 
\end{table*}
\end{center}


\section{{\sc ProDiMo} code description}\label{prodimo}

{\sc ProDiMo} is designed to compute self-consistently the (1+1)D
  hydrostatic disc structure in thermal balance and kinetic chemical
  equilibrium. The disc can be active or passive depending on the
choice of the viscosity parameter $\alpha$.  We first provide here a
brief description of {\sc ProDiMo}. Interested readers are referred to
\citet*{Woitke2009A&A...501..383W} to find detailed explanations of
the physics implemented in the code. We continue by giving the
rational for the ``soft-inner edge'' surface density profile
implemented in {\sc ProDiMo}. We finish by explaining the computation
of Spectral Energy Distributions and images, which is a new feature of
the code.

\subsection{General description}

The code iterates the computation of the disc density, gas and dust
temperature, and chemical abundance structure until successive
iterations show less than 1\% change.

The iteration starts with a given hydrostatic disc structure computed
from the previous iteration or from an assumed structure. The dust
radiative transfer module computes the dust temperatures and the disc
local mean specific intensities $J_\nu(r,z)$. The specific intensities
are used to calculate the photochemical rates. The 2D dust
radiative-transfer module of {\sc ProDiMo} has been benchmarked
against other codes that use other methods such as Monte-Carlo
\citep{Pinte2009A&A...498..967P}. The continuum radiative transfer
includes absorption, isotropic scattering, and thermal
  emission. The grains are assumed spherical. The lack of anisotropic
scattering prevents us to calculate completely accurate images. In
case of anisotropic scattering, one expects the far upper part of the
disc rim to be dimmer (backward scattering) than the near lower part
(forward scattering). However, isotropic scattering is sufficiently
accurate for generating precise SEDs.

The grains have sizes between $a_{\mathrm{min}}$ and  
$a_{\mathrm{max}}$ and follow a power-law size distribution with index
$p$. The absorption and scattering efficiencies are computed using Mie
theory for compact spherical grains.  Grains of different sizes are
assumed to have the same temperature. This limitation in our model
does not prevent the comparison between self-consistently computed gas
temperature models (thermal-decoupled models) and models with
$T_{\mathrm{gas}}=T_{\mathrm{dust}}$ (thermo-coupled models). In this
paper, the dust composition and size distribution are constant
throughout the disc. We adopt the interstellar dust optical constants
of \citet{Laor1993ApJ...402..441L} for amorphous grains since we are
only interested in modelling featureless broad SEDs.

Once the dust temperature $T_{\mathrm{dust}}(r,z)$ and the continuum
mean specific intensities $J_\nu(r,z)$ are known, the gas temperature
and chemical concentrations are consistently calculated assuming
thermal balance and kinetic chemical equilibrium. The chemistry
network includes 71 gas and ice species. The chemical reactions
comprise photo-reactions (photoionization and photodissociation) with
rates computed using cross-sections and specific intensities $J_\nu$
calculated by the 2D radiative transfer module, gas phase reactions,
gas freeze-out and evaporation (thermal desorption, photodesorption,
and cosmic-ray induced desorption), and H$_2$ formation on grain
surfaces. The chemical rates are drawn from the {\sc UMIST} 2007
database \citep{Wooddall2007A&A...466.1197W} augmented by rates from
the {\sc NIST} chemical kinetic website. The photo-cross-sections are
taken from \citet{vanDishoeck2008}. The H$_2$ formation
on grain surfaces follows the prescription of
\citet{Cazaux2004ApJ...604..222C} with the most recent grain surface
parameters.
  
The gas can be heated by line absorption and cooled by line emission.
The line radiative transfer is computed via the escape probability
formalism. The gas is also heated by interactions with hot
photoelectrons from dust grains and Polycyclic Aromatic Hydrocarbons
and with cosmic rays.  Thermal accommodation on grain surfaces can
either heat or cool the gas depending on the temperature difference
between the gas and the dust
\citep{Tielens2005pcim.book.....T}. Thermal accommodation dominates at
high densities and optical depths, driving the gas and dust towards
the same temperatures.

Finally, the gas temperature and molecular weight determine the local
pressure at each grid point, which is used to modify the disc structure
according to the vertical hydrostatic equilibrium. The new disc structure is
compared to the previous one to check for convergence.


\subsection{Soft Inner edge}  

The radial dependence of the vertical column density is assumed to
deviate from a power-law in the innermost layers, and is implemented
according to the "soft-edge"-description of \citealp*[(see Section
3.1)]{Woitke2009A&A...501..383W}. We assume that the gas has been pushed
inwards by the radial pressure gradients around the inner edge, and
has spun up according to angular momentum conservation, until the
increased centrifugal force balances the radial pressure gradient. The
procedure results in a surface layer in radial hydrostatic equilibrium
at constant specific angular momentum where the column density
increases gradually from virtually zero to the desired value at the
point where we start to apply the power-law. The thickness of this
"soft" edge results to be typically a few 10\% of the inner
radius $R_{\mathrm{in}}$.
\begin{figure}
\centering
\resizebox{\hsize}{!}
{\includegraphics[width=18cm]{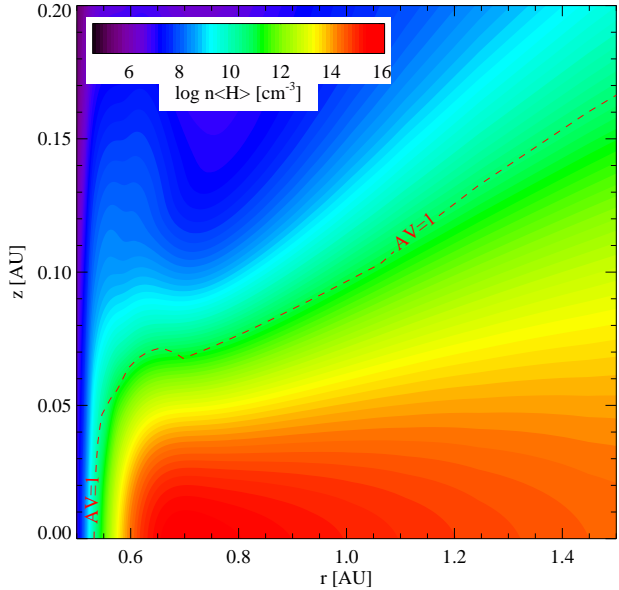}}
\caption{Inner rim density structure and photospheric height
  ($A_{\mathrm{V}}$=1 contour) for the fiducial model (``soft-edge''
  and $T_{\mathrm{gas}}$ computed by thermal balance).}
  \label{inner_rim_photospheric_height}
\end{figure}
\subsection{Spectral Energy Distribution calculation}
\begin{figure}
\centering
\resizebox{\hsize}{!}{\includegraphics[]{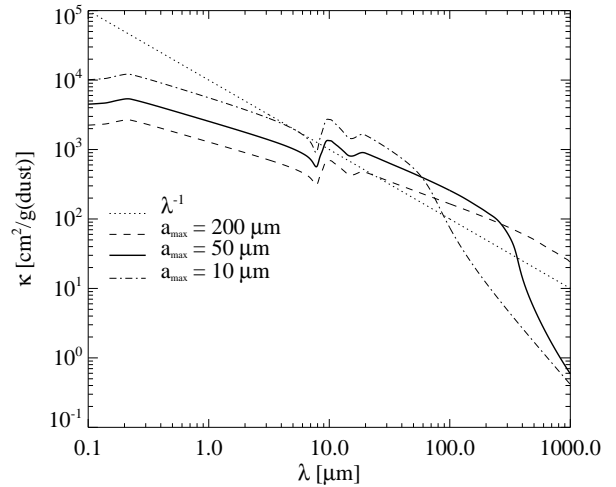}}
\caption{ The dust opacity computed by Mie theory for the three values
  of $a_{\mathrm{max}}$ (10, 50, 200 $\mu$m) and a power-law dust grain
  size distribution (index 3.5). A line showing an opacity with a
  $\lambda^{-1}$ opacity law is added for comparison.}
  \label{fig_opacity} 
\end{figure}

Based on the results of the continuum dust radiative transfer solution
(see \citet{Woitke2009A&A...501..383W}, Sect.\ 4) we have developed a
new {\sc ProDiMo} module for SED and image calculation based on formal
solutions of the dust continuum radiative transfer equation along a
bundle of parallel rays that cross the disc at given inclination angle
$i$ with respect to the disc rotation axis.  The rays start at an
image plane put safely outside of the disc. The ray direction
$\overrightarrow{n}$ and the 3D origin of the image plane
$\overrightarrow{p_0}$ are given by
\begin{equation}
\overrightarrow{n}=\left(\begin{array}{c}
\sin i\\
0\\
\cos i\end{array}\right)
\end{equation}
and
\begin{equation}
\overrightarrow{p_{0}}=R\ \overrightarrow{n},
\end{equation}
where $R$ is a large enough distance to be sure that the plane cannot
intersect the disc. The plane is aligned along the two perpendicular
unit vectors
\begin{equation}
\overrightarrow{n_{z}}=\left(\begin{array}{c}
0\\
1\\
0\end{array}\right)
\end{equation}
and
\begin{equation}
\overrightarrow{n_{y}}=\left(\begin{array}{c}
-\cos i\\
0\\
\sin i\end{array}\right).  
\end{equation}
The rays are organized in log-equidistant concentric rings in the
image plane, using polar image coordinates ($r$,$\theta$). The 3D
starting point of one ray $\overrightarrow{x_0}$, with image
coordinates $x=r \sin(\theta)$ and $y=r \cos(\theta)$, is given by
\begin{equation}
\overrightarrow{x_{0}}=\overrightarrow{p_{0}}+x\ \overrightarrow{n_{x}}+y\ \overrightarrow{n_{y}}
\end{equation}
From these ray starting points $\vec{x}_0$, the radiative transfer
equation is solved backwards along direction $-\overrightarrow{n}$,
using an error-controlled ray integration scheme as described in
Woitke, Kamp, \& Thi (2009, Sect 4.2). The source function and opacity
structure in the disc have been saved from the previous run of {\sc
  ProDiMo}'s main dust radiative transfer. The resulting intensities
at the location of the image plane, $I_\nu(r,\theta)$, are used to
directly obtain monochromatic images.

\begin{figure*}
\centering  
  {\includegraphics[angle=0,width=18cm,height=19cm]{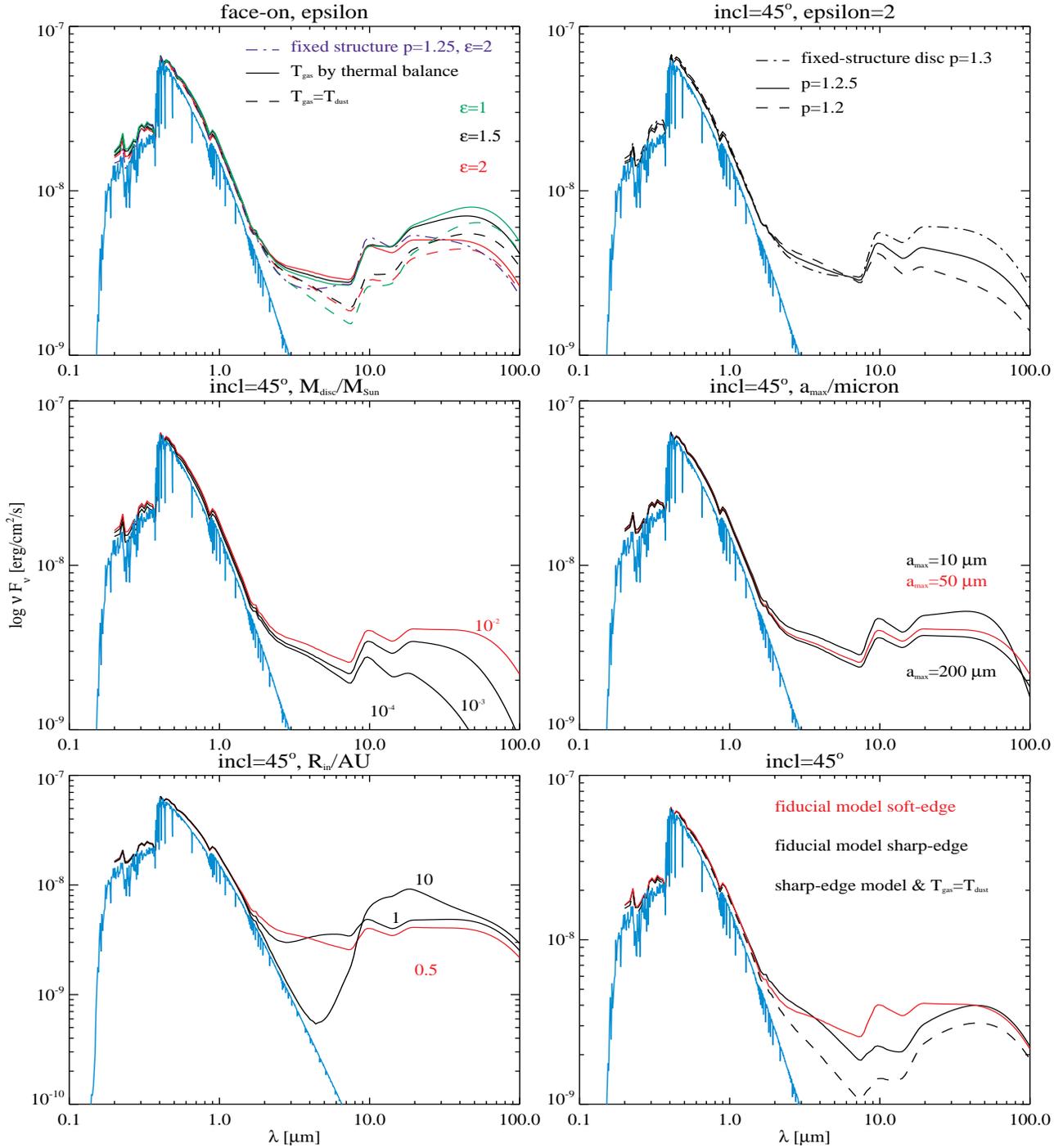}}
  \caption{Spectral energy distribution computed for the various
    models. The upper left panel shows the fiducial model
    ($M_{\mathrm{disc}}$=10$^{-2}$ M$_\odot$) with decreasing surface
    power index $\epsilon$ and in dot-dash purple line the SEDs
    computed for the fixed disc structure models with the flaring
    index $p$=1.25 and $\epsilon$=2. The upper-right panel shows the
    SED for the fixed-structure model with $\epsilon=2$ and three
    values of the flaring index $p$. The upper-left panel is for discs
    seen face-on. All the other panels are for discs seen at 45
    degree. The other panels show the effect of varying the disc total
    mass (10$^{-2}$, 10$^{-3}$, and 10$^{-4}$ M$_\odot$), the grain
    upper size limit ($a_{\mathrm{max}}=$~10~$\mu$m and
    $a_{\mathrm{max}}=$~200~$\mu$m), the inner disc radius (0.5, 1,
    and 10~AU), and the effect of adopting a ``sharp-edge'' density
    profile compared to a ``soft-edge'' density profile. SEDs in solid
    lines are computed in the case $T_{\mathrm{gas}}$ is computed by
    thermal balance while SEDs in dash lines are computed assuming
    $T_{\mathrm{gas}}=T_{\mathrm{dust}}$. The SED of the fiducial
    model is shown in red. The input stellar spectrum is plotted in
    light blue. The spectral sampling is much higher for the stellar
    spectrum than for the SEDs. The flux in the UV and optical is the
    sum of the direct stellar flux and the stellar flux scattered by
    the disc.}
  \label{fig_SEDs}
\end{figure*}

To retrieve the spectral flux under inclination $i$, we integrate over
the image plane as
\begin{equation}
 F_\nu = \frac{1}{d^2} \int I_\nu(r,\theta) r dr d\theta
\end{equation}
where $d$ is the object's distance. We use $N_\theta=160$ and $N_r=200$
rays. Due to the log-spacing of the concentric rings, the method
safely resolves even tiny structures originating from the inner rim.

\begin{figure*}
\centering
  {\includegraphics[width=18cm]{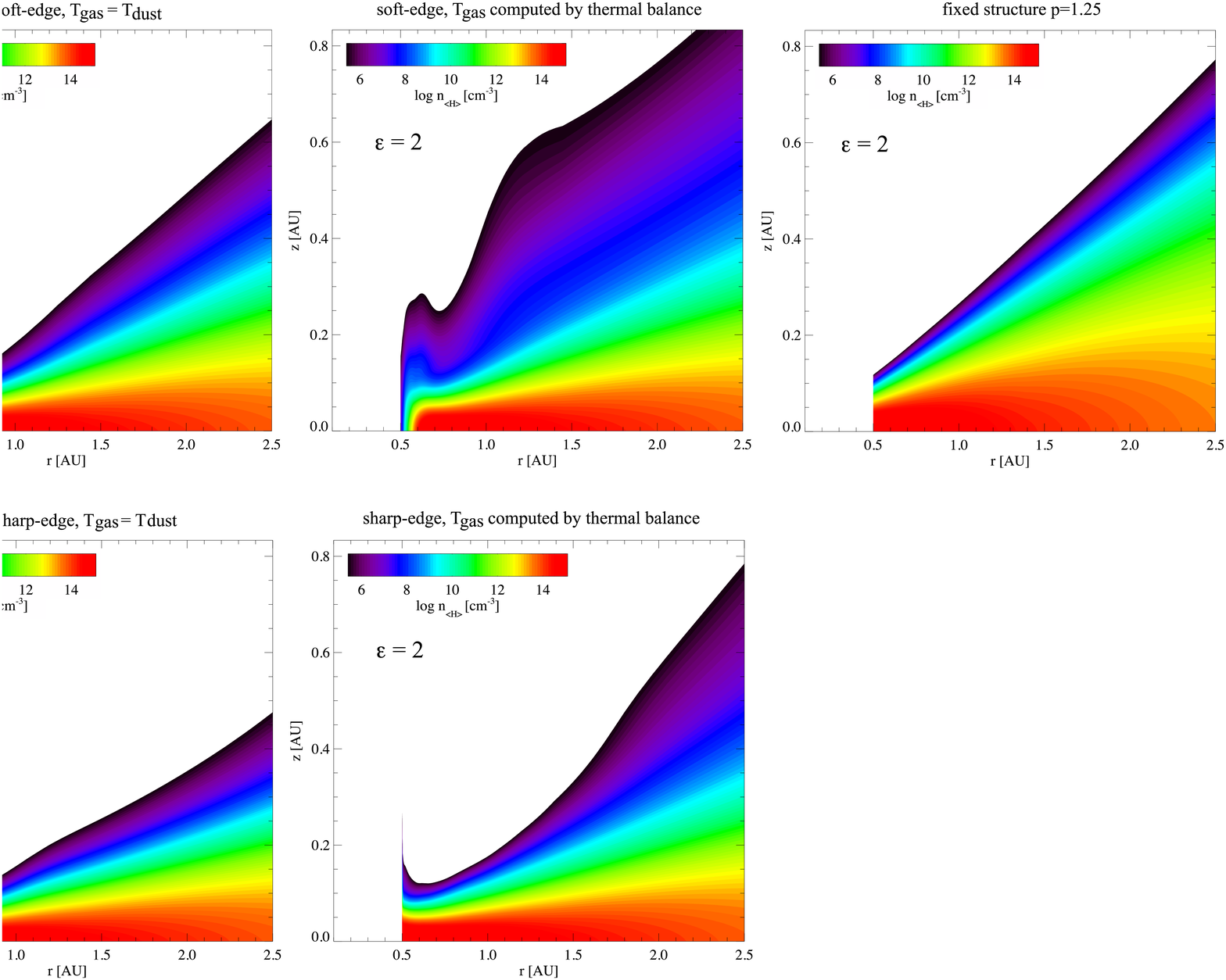}}
  \caption{Inner rim density structure for the fiducial model but with
    different disc structure prescriptions as indicated above each
    panel.}
  \label{inner_rim_structures}
\end{figure*}

\begin{figure*}
\centering   
  {\includegraphics[width=18cm]{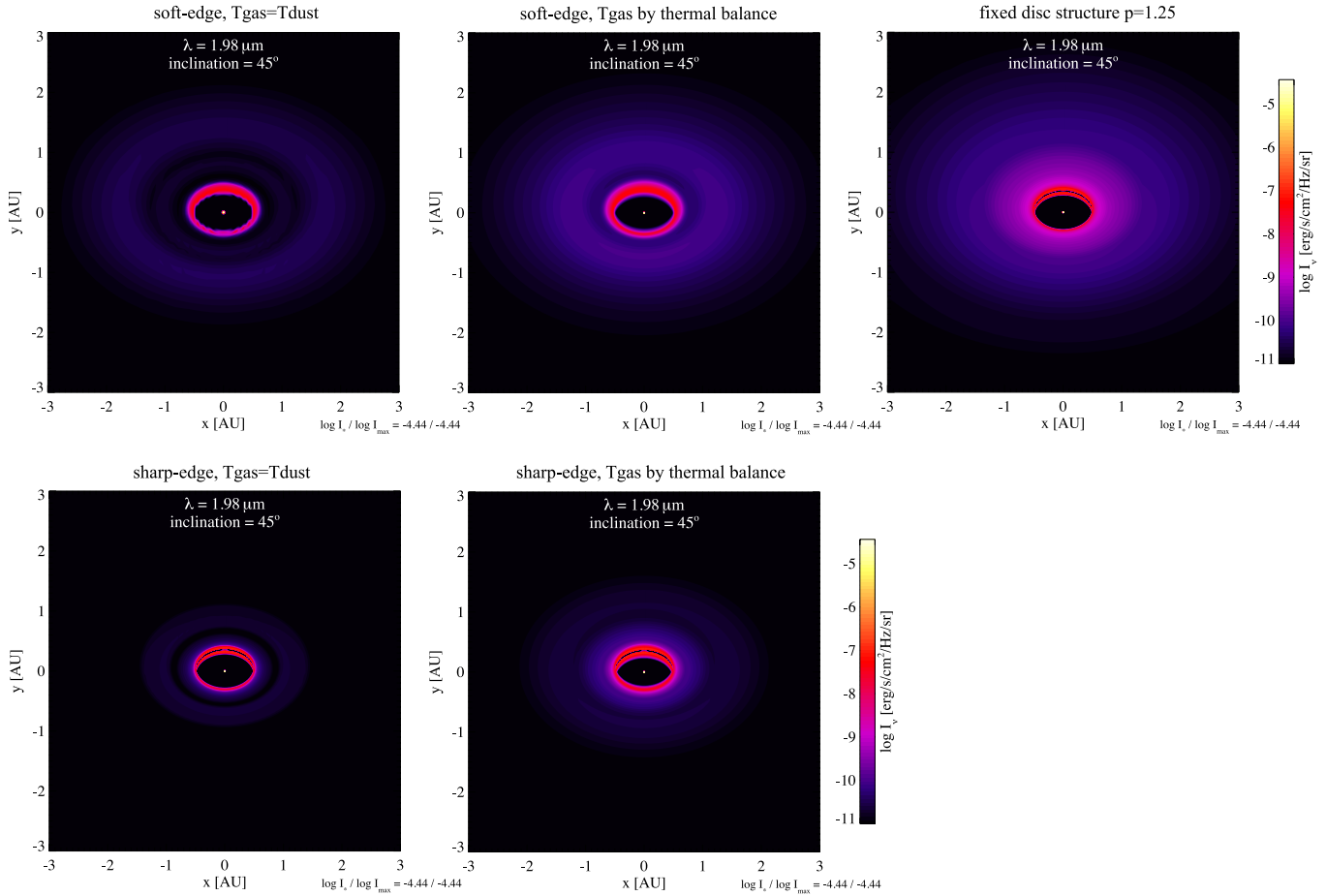}}
  \caption{Image of the inner disc at 1.98 $\mu$m for the fiducial
    model ($M_{\mathrm{disc}}$=10$^{-2}$ M$_\odot$, $\epsilon=2$) for
    the two possible combinations of inner edge prescriptions and two
    ways to compute the gas temperature and for the fixed-structure
    disc structure. In the fixed structure case, the scale-height is
    0.72~AU at radius 10~AU and the flaring index is 1.25.  The disc
    is seen at an inclination of 45 degree. The inner radius was set
    at 0.5~AU. We can see a second rim at 1--2~AU, which is the most
    pronounced in the case of soft-edge and gas temperature computed
    by thermal balance. The emission from the second bump contributes
    significantly to the total flux. The fixed-disc model has a
    flaring index of $p=$~1.25.}
  \label{fig_image_1.98_rim}
\end{figure*}
   
\begin{figure*}
\centering 
  {\includegraphics[width=18cm]{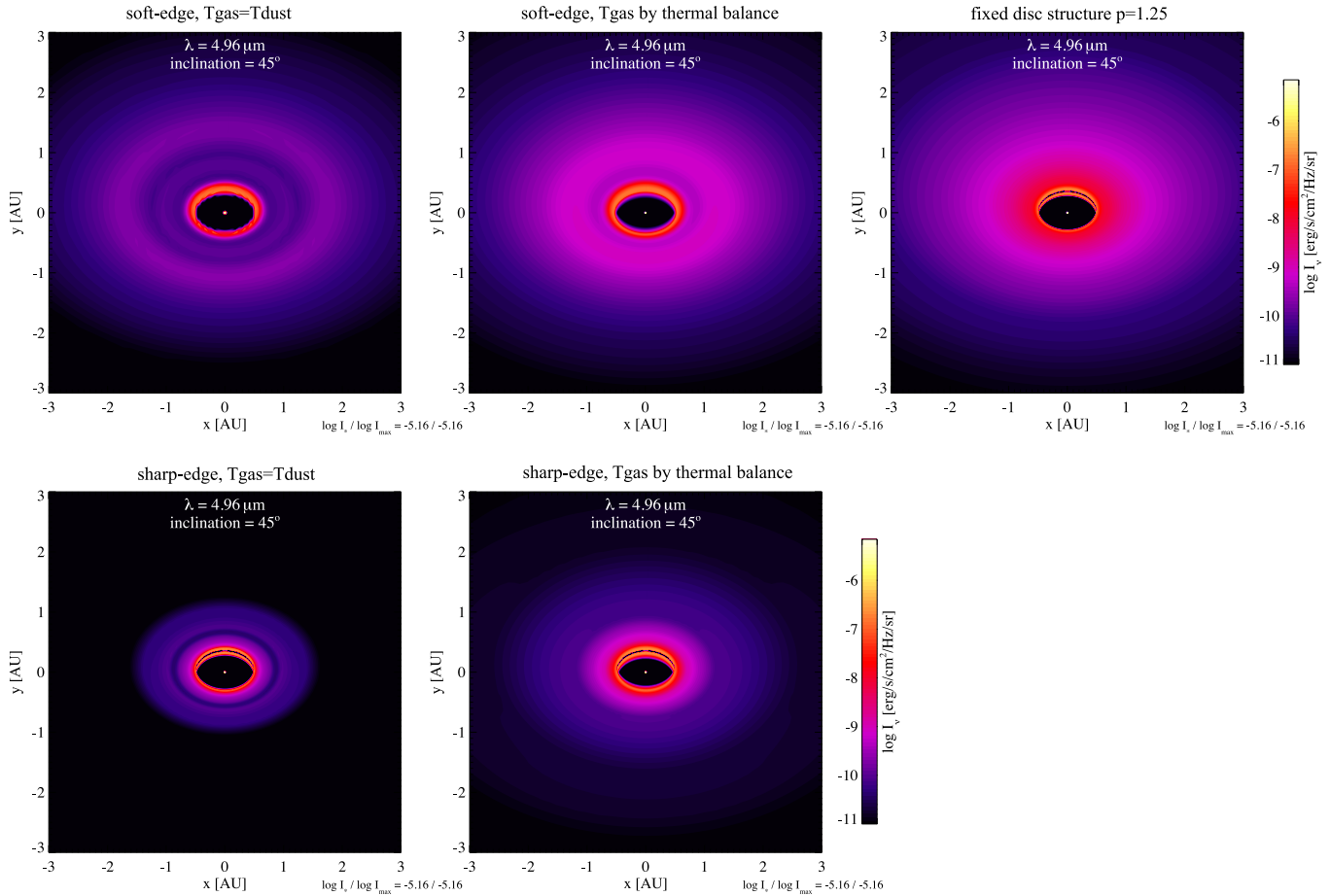}}
  \caption{Same as Fig.~\ref{fig_image_1.98_rim} but the images were
    generated at 4.96 $\mu$m}
  \label{fig_image_4.96_rim}
\end{figure*}
  
\section{Disc parameters}\label{model_description}

\subsection{Fiducial model}\label{fiducial_model_description}
\begin{figure*}
\centering 
  {\includegraphics[width=18cm,angle=0]{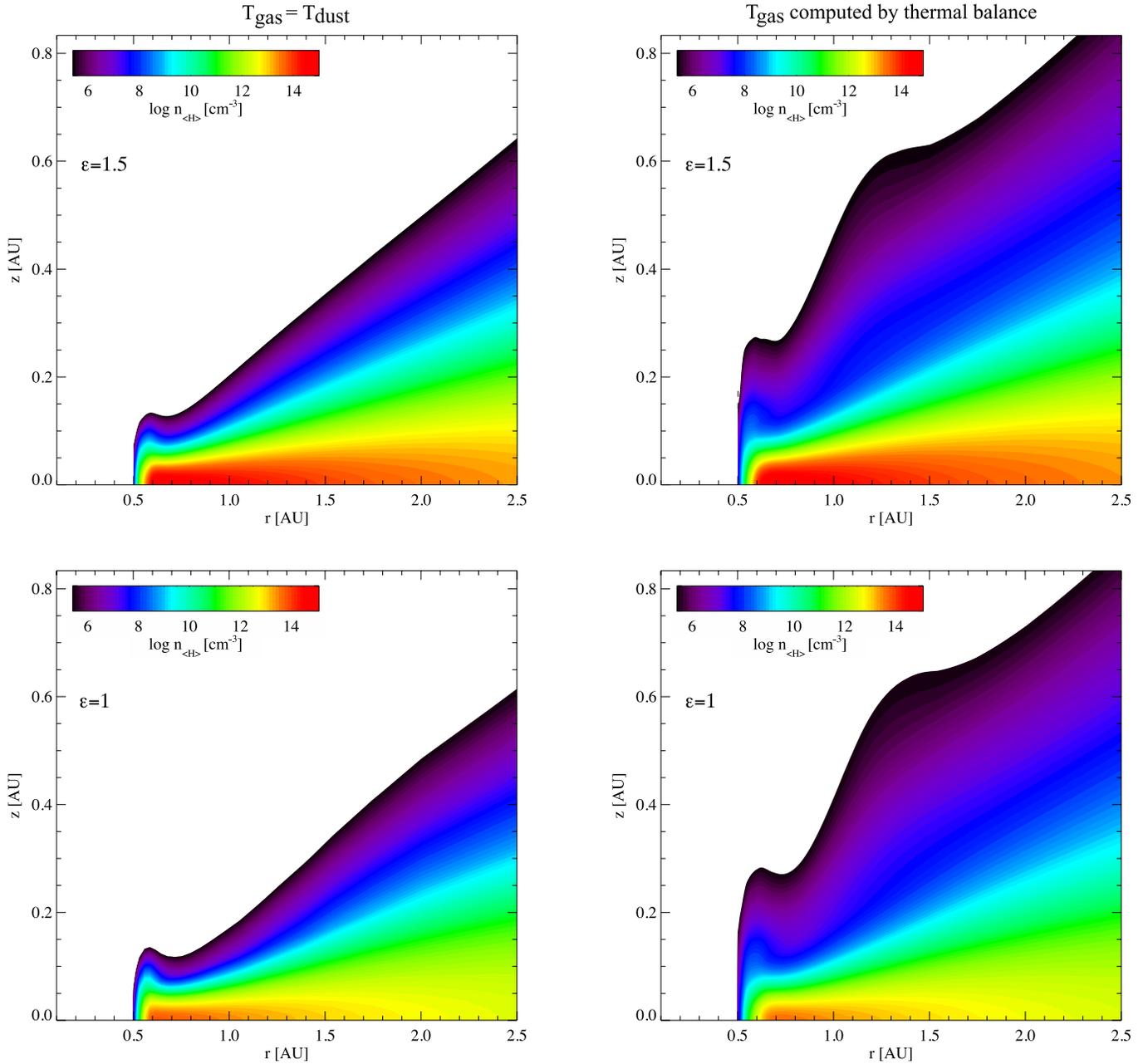}}
  \caption{Inner rim density structure for the fiducial model
    ($M_{\mathrm{disc}}$=10$^{-2}$ M$_\odot$) with surface power index
    $\epsilon=$1.5 and $\epsilon=$1. The left panels are the models
    with $T_{\mathrm{gas}}=T_{\mathrm{dust}}$. The right panels are
    the models with $T_{\mathrm{gas}}$ computed by gas thermal
    balance. These models show a much taller rim and also a secondary
    rim at $\sim$~1.2 AU.}
  \label{fig_surf_density_inner_rim}
\end{figure*}


\begin{figure*}   
\begin{minipage}[b]{0.48\linewidth}
\centering
{\includegraphics[scale=0.48]{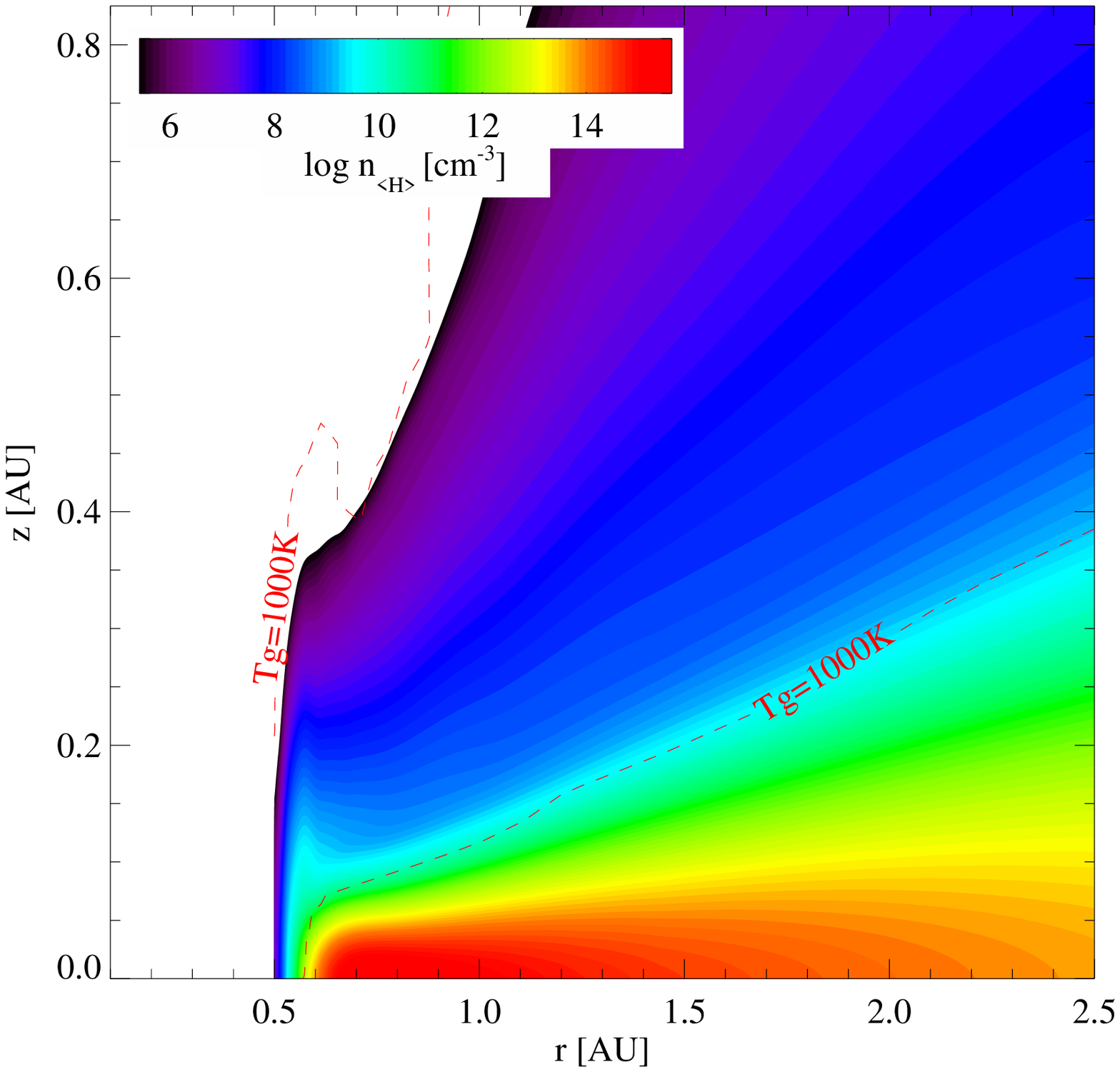}}
\end{minipage}
\begin{minipage}[b]{0.48\linewidth}
\centering
{\includegraphics[scale=0.48]{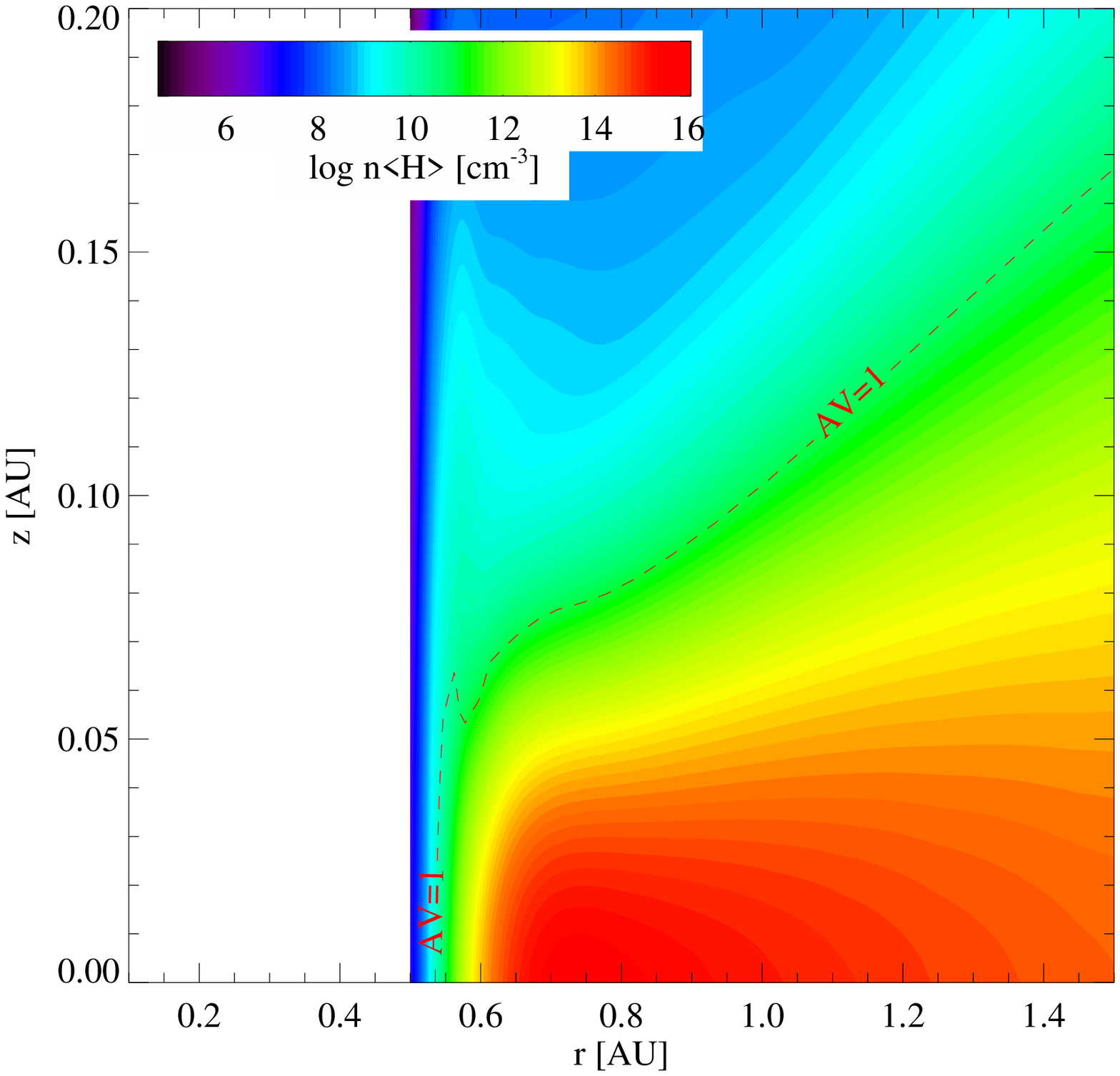}}
\end{minipage}
\begin{minipage}[b]{0.48\linewidth}
\centering
{\includegraphics[scale=0.48]{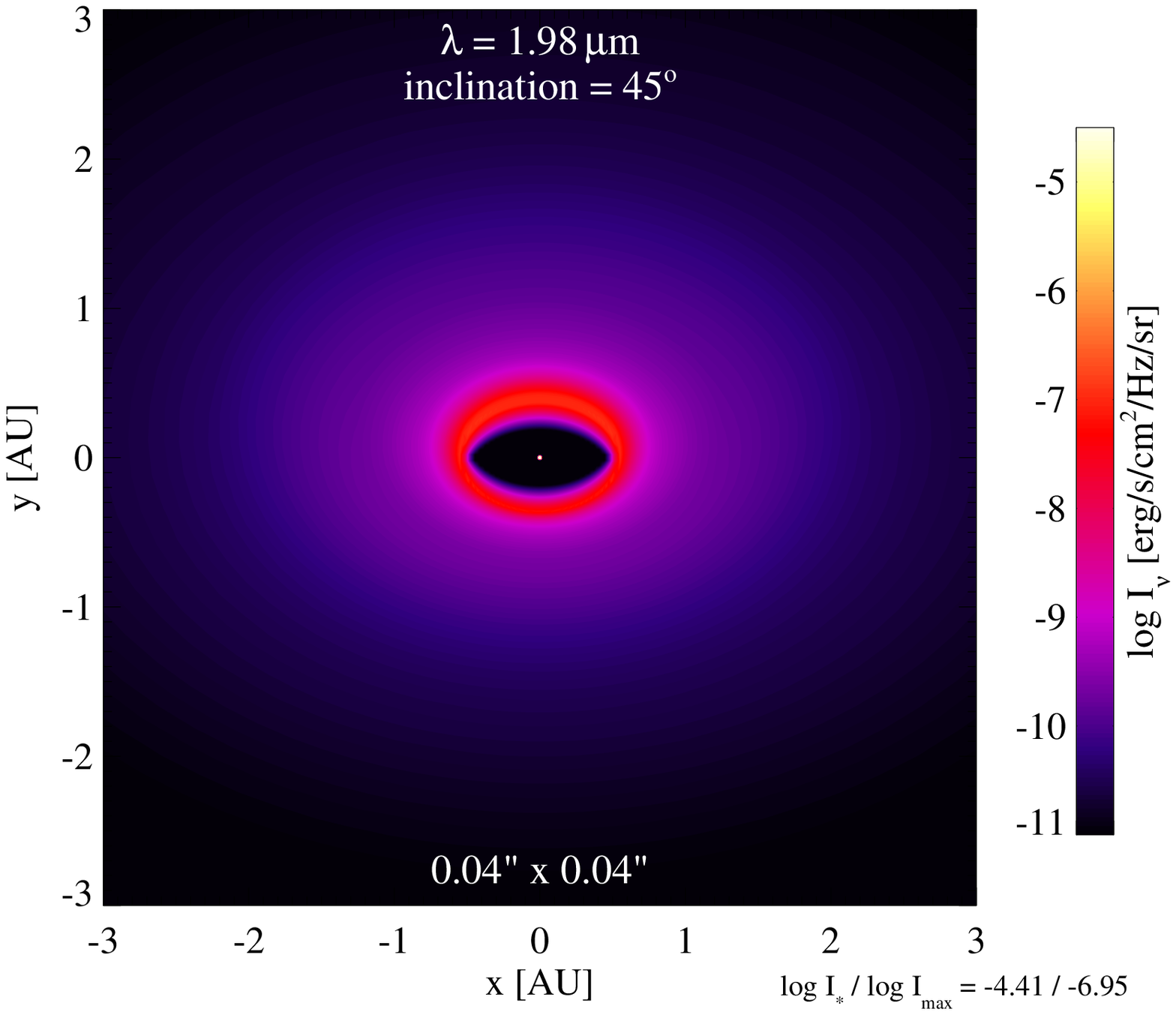}}
\end{minipage}
\begin{minipage}[b]{0.48\linewidth}
\centering
{\includegraphics[scale=0.48]{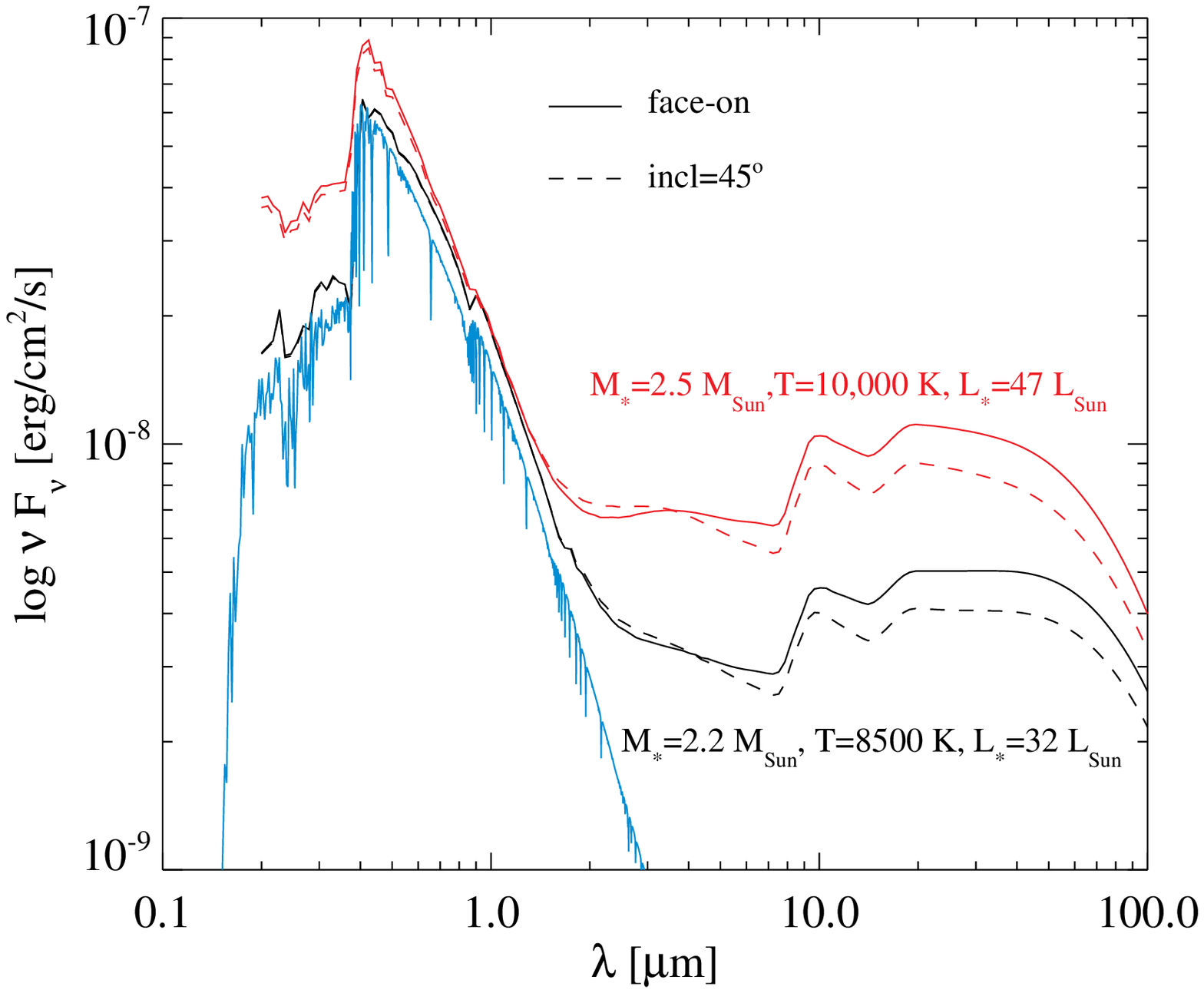}}
\end{minipage}
\caption{\label{fig_DDN} Inner disc density structure (upper-left
  panel), photospheric height (upper-right panel), near-infrared image
  (lower-left panel), and comparison between the SED for the fiducial
  model (lower-right panel) but with M$_*$=2.5 M$_\odot$,
  $T_{\mathrm{eff}}$=10,000~K, and $L_*$=47 L$_\odot$. We show the
  $L_*=$~32~$L_\odot$ stellar spectrum only. The red-dashed line in the
  upper-left panel encloses the region where the gas temperature is
  higher than 1000~K. Notice the large amount of scattered light in
  the blue part of the visible range in the SEDs caused by the
  extremely flaring inner disc atmosphere for the more luminous star.}
\end{figure*}

We modelled circumstellar discs around one typical Herbig~Ae star with
effective temperature of 8600~K, mass of 2.2~M$_{\odot}$, and
luminosity of 32~L$_{\odot}$. In this study, we did not try to fit the
SED of any particular object. The input stellar spectrum is taken from
the {\sc PHOENIX} database of stellar spectra
\citep{Brott2005ESASP.576..565B}.

The discs are illuminated from all sides by a low interstellar medium
UV flux ($\chi=0.1$) in addition to the stellar flux. We chose a low
flux because previous studies do not take interstellar UV flux into
account. The stellar flux dominates over the standard interstellar
flux in the inner disc by orders of magnitude.

The abundance of PAH ($f_{\mathrm{PAH}}$) in discs controls the gas
temperature and the disc flaring, which in turn affects the dust
temperature. The exact value of $f_{\mathrm{PAH}}$ can only be constrained by
a simultaneous fit to the broad SED and the PAH features for a specific 
object. We chose an arbitrary low abundance (1\%) and tested the effects on
the SED for the fiducial model with $f_{\mathrm{PAH}}$=0.1.

The only parameters that were allowed to change are the disc total
mass ($M_{\mathrm{disc}}=$~10$^{-4}$, 10$^{-3}$, and 10$^{-2}$
M$_\odot$), the surface power-law index ($\Sigma=\Sigma_o \times
(r/r_o)^{-\epsilon}$ with $\epsilon=$~1.0, 1.5, and 2.0), the disc
inner radius ($R_{\mathrm{in}}=$~0.5, 1 , 10~AU), and the maximum
grain radius ($a_{\mathrm{max}}=$~10, 50, 200~$\mu$m). 

We chose $\epsilon=2$ as our fiducial value for the surface density
profile index. Surface density profiles as steep as $r^{-2}$ for the
inner disc are a signature of self-gravitating discs
\citep{Rice2009MNRAS.396.2228R}. However, we consider discs with
maximum disc mass over star mass ratio of 4.5 $\times$ 10$^{-3}$. A
recent derivation of the Solar Nebula also gives a steep surface
density profile inside 50~AU \citep{Desch2007ApJ...671..878D}.
Shallower decreases in surface density ($\epsilon=1.5$) are more
commonly observed for the outer disc ($R>$~50~AU).

The density falls off at $R_{\mathrm{in}}$ following the ``soft-edge''
model described earlier.  The minimum grain size $a_{\mathrm{min}}$ is
kept at 0.05 $\mu$m. The power-law index of the dust grain size
distribution follows the standard interstellar value of -3.5. A
power-law index of -3.5 results from grain-grain collision theory
\citep{Dohnanyi1969JGR....74.2531D}. We mimic grain growth by varying
the maximum grain radius $a_{\mathrm{max}}$ from the fiducial value of
50 to 200 $\mu$m. We also studied the effect of smaller grains by
setting the maximum grain size to 10~$\mu$m. 

The choice of the inner disc radius at 0.5 AU was dictated by disc
accretion and dust grain physics. A disc is truncated at the
co-rotation radius in the absence of a magnetic field or of a low-mass
companion, either star, brown dwarf, or giant planet, orbiting close
to the star. The co-rotation radius
$r_{\mathrm{rot}}=GM_*/v_{\mathrm{rot}}^2$ is defined as the distance
from the central star where the Keplerian disc rotates at the same
speed than the star photosphere $v_{\mathrm{rot}}=v_*$ (e.g.,
\citealt{Shu1994ApJ...429..781S}). Assuming a photospheric rotation
speed $v_* = 50-150$ km s$^{-1}$, typical of young stars
\citep{vandenAncker1998A&A...330..145V}, we obtain
$r_{\mathrm{rot}}\simeq$ 0.08-0.76 AU for $M_*=2.2$ M$_\odot$.  Dust
grains do not exist above their sublimation temperature, which is
around 1500~K for silicate grains. The dust sublimation radius is the
distance from the star beyond which dust exists and can be estimated by
$r_{\mathrm{d}}=\sqrt{Q_R(L_*+L_{\mathrm{acc}})/16 \pi
  \sigma}/T_{\mathrm{sub}}^{2}$ where
$Q_R=Q_{\mathrm{abs}}(a,T_*)/Q_{\mathrm{abs}}(a,T_{\mathrm{sub}})$ the
ratio of the dust absorption efficiency at stellar temperature $T_*$
to its emission efficiency at the dust sublimation temperature
$T_{\mathrm{sub}}$, $a$ is the mean grain radius and $\sigma$ is the
Stefan constant. For large silicate grains ($a \geq 1\ \mu$m), $Q_R$
is relatively insensitive to the stellar effective temperature and
close to unity because most stellar radiation lies at wavelengths
shorter than the grain radius. For smaller grains, the value of $Q_R$
is significantly increased \citep{Monnier2002ApJ...579..694M}.  For
one micron grains, the sublimation radius in our fiducial model is
$r_{\mathrm{d}}= 0.19$~AU assuming $L_{\mathrm{acc}}<<L_*$ (passive
disc) and $Q_R \simeq 1$ (large grains). For grains as small as 0.05
$\mu$m in radius, we obtain $Q_R \simeq 20$ and $r_{\mathrm{d}}=
0.88$~AU. The inner disc location at 0.5~AU in our models is within
the theoretical range of inner disc radii. Moreover grains as small as
0.05$\mu$m can survive at 0.5~AU. Large luminosity ($L_*>100$
L$_\odot$) fast rotators ($v \sin i>200$ km s$^{-1}$) like 51~Oph
(with $L_*$= 260$^{+60}_{-50}$ L$_\odot$ and $v_*$=267$\pm$5 km
s$^{-1}$ ) may have a inner dust-free gas-rich region between the
co-rotation radius and the dust sublimation radius
\citep{Thi2005A&A...430L..61T,Tatulli2008A&A...489.1151T}. Another
example is AB Aur where the corotation radius is smaller than the
sublimation radius for the small grains but not for the large
grains. Observational evidence of an inner gas-rich dust-poor region
has been found for this object \citep{Tannikurlam2008ApJ...677L..51T}.

We also chose to simulate discs with inner radius at 1 and 10~AU to
model typical gaps possibly created by planets. The outer radius
$R_{\mathrm{out}}$ was set at 300~AU. By default passive discs are
modelled (i.e., $\alpha=$~0). Since we focus on young discs the
standard interstellar medium value of 100 for the gas-to-dust mass
ratio was adopted for all models. The gas and dust were assumed
well-mixed with no dust settling. The common model parameters are
summarized in Table~\ref{tab_DiscParameters} with the parameters for
the fiducial model printed in boldface .  For each parameter set, the
code was run in the thermal-coupled mode
($T_{\mathrm{dust}}=T_{\mathrm{gas}}$) and thermal-decoupled mode in a
50 $\times$ 50 non-regular grid. An example of the location of the
grid points is given by \citet{Kamp2010A&A...510A..18K}. All SEDs were
computed for a source located at the typical distance of 140~pc at two
inclinations: face-on (0 degree) and 45 degree.

The stellar properties of our fiducial model differs from earlier
studies ($T_{\mathrm{eff}}$= 10000~K, $M_*$=2.5 M$_\odot$, and
$L_*$=47~L$_\odot$). We plot in Fig.~\ref{fig_DDN} the density
structure, photospheric height, near-infrared image, and SEDs for two
inclinations using the earlier studies properties. The disc structures
and SEDs are different for the two sets of stellar parameters. The
hotter disc atmosphere around the 47 L$_\odot$ star flares much more
than the disc around the lower luminosity star. The infrared
luminosity is increased but the shape of the SED does not differ much
as the stellar luminosity is increased.

\subsection{Fixed-structure model}\label{mcfost_model_descrption}
\begin{center}
\begin{table}
  \caption{Disc parameters for the fixed-structure model.}\label{tab_MCFOSTParameters}
		\begin{tabular}{llll}
                  \hline
                  reference scale height & $H_0$ & 0.72 & AU\\
                  reference radius       & $R_0$ & 10   & AU\\
                  flaring index          & $p$   & 1.2, {\bf 1.25}, 1.3 & \\
                  \hline
\end{tabular}
\ \\ 
\end{table}
\end{center}
We compare the results of the hydrostatic disc models to the results
of a fixed-structure model. The disc height $H$ is parametrised by the
functional $H(R)=H_0(R/R_0)^p$, where $H_0$ is the reference scale
height at reference radius $R_0$ and $p$ is the flaring index. The
scale height is used to compute the density $n_{\mathrm{H}}(R,z)$ at
height $z$ by $n_{\mathrm{H}}(R,z)=n_0(R)exp(-(z/H)^2)$. The value of
the parameters are given in Table~\ref{tab_MCFOSTParameters} and were
chosen such that the SED matches that produced by the fiducial
  model with ``soft-edge'' and $T_{\mathrm{gas}}$ computed by thermal
  balance. We vary the value of the flaring index $p$ between 1.2 and
1.3 to study its effects on the SEDs. All the other parameters are the
same than for the fiducial model.

\section{Results and discussion}\label{results_discussion}

In this section, we discuss the effects on the disc gas density
structure and SED when we vary a few parameters in the case of equal
gas and dust temperature and when they are computed independently. We
first show the structure and SED of the fiducial disc in
Sect.~\ref{fiducial_model}. The SEDs and images of the fixed-structure
disc is discussed in Sect.~\ref{mcfost_model_results}. We continue by
addressing the effects on the disc structure and SED when we use the
``soft-edge'' versus the ``sharp-edge'' inner rim prescription
(Sect.~\ref{sharp_soft_edge}), when we vary the surface density
profile (Sect.~\ref{surf_density_profile}), the disc mass
(Sect.~\ref{disc_mass}), the maximum grain size
(Sect.~\ref{big_grains}), and the inner disc radius
(Sect.~\ref{inner_radius}).

\subsection{Fiducial model}\label{fiducial_model}

Figure~\ref{fig_fiducial} contains some results for the fiducial
model.  The dust and gas temperature distribution are shown in the
upper two panels while the lower-right panel is the disc density
structure. The disc areas where the gas and dust temperature are
decoupled are emphasized in the lower-left panel.
Figure~\ref{fig_opacity} shows the dust opacity (scattering and
absorption) for three maximum grain sizes (10, 50, and 200~$\mu$m).
The opacity is dominated by scattering at wavelengths below 10~$\mu$m
and by absorption at longer wavelengths.

In the thermal-coupled models, the dust temperature structure is
divided into a hot inner rim, a warm upper layer, and a vertical
isothermal interior. The lower-left panel in Fig.~\ref{fig_fiducial}
shows the ratio between the gas and dust temperature. The gas and dust
temperature are equal at vertical optical depths $A_{\mathrm{V}}>$~1
for a 10$^{-2}$ M$_\odot$ disc. The 2D dust continuum
radiative-transfer results confirm the analytical two-zone
decomposition popularized by \citet{Chiang1997ApJ...490..368C}.  The
SED of the fiducial model is shown in solid-red lines in
Figure~\ref{fig_SEDs}. Model images of the inner disc at 1.98 $\mu$m
and 4.96~$\mu$m are shown in Fig.~\ref{fig_image_1.98_rim} and
Fig.~\ref{fig_image_4.96_rim} respectively. The flux at 1.98~$\mu$m
comes mostly from the inner rounded rim. The region just behind the
rim is deprived of stellar photons, is cooler, and thus does not emit
strongly (shadowing effect).

The gas temperature remains above 1000~K in the disc atmosphere up to
a few AU. Together with the decrease in the vertical gravitational
pull, they explain the emergence of a secondary density bump at
$\sim$~1.3~AU. The presence of the hot ''finger'' in disc atmospheres
is typical of disc models that compute the gas and dust temperature
independently
\citep*{Ercolano2009ApJ...699.1639E,Glassgold2009ApJ...701..142G,Woitke2009A&A...501..383W,Nomura2005A&A...438..923N,Kamp2004ApJ...615..991K,Jonkheid2004A&A...428..511J}.

  The shape of the disc photospheric height ($A_{\mathrm{V}}$=1
  contour) is shown in Figure~\ref{inner_rim_photospheric_height}. The
  height at 0.6 AU is located at $z/r \sim$~1. The curvature exhibits
  the same shape than in models where grains of different sizes
  sublime at different radii \citep{Isella2005A&A...438..899I} or
  with density-dependent sublimation temperature
  \citep{Tannirkulam2007ApJ...661..374T}. The curvature of the inner
  rim is caused by gas angular momentum conservation and pressure
  gradient and is independent on the dust physics. Likewise the
  location of the inner rim is set by hydrodynamic constraints (disc
  truncation or presence of a companion) and not by dust sublimation
  physics (see Sect.~\ref{fiducial_model_description}).  However we
  have assumed that the gas and the dust are well-mixed. Therefore,
  the gas and dust grain density structures may be independent if
  grain-growth and settling occur.

\subsection{Fixed-structure model}\label{mcfost_model_results}
  
The inner disc structure for the $H_0=0.72$~AU and flaring index
$p=$1.25 fixed-structure model is shown in the upper-right panel of
Fig.~\ref{inner_rim_structures}.  The inner rim height is
$\sim$~0.1~AU. The SEDs that correspond to the fixed-structure models
are drawn for inclination 0 ($p$=1.25) and 45 degree ($p$=1.2, 1.25,
1.3) on the two top panels of Fig.~\ref{fig_SEDs}. The flaring index
  $p$ impacts on the amount of warm dust grains in the disc atmosphere
  that emit beyond 100~$\mu$m.  The SED of the fixed-structure model
with flaring index $p=$1.25 matches relatively well the SED of the
fiducial model. The images at 1.98 and 4.96~$\mu$m for the $p=$1.25
model are shown on the upper right panel of
Fig.~\ref{fig_image_1.98_rim} and Fig.~\ref{fig_image_4.96_rim}.
  
When viewed exactly face-on, the SEDs at 3$\mu$m of a fixed-structure
disc show very weak near-IR fluxes, which arise from the rim
\citep{Meijer2008A&A...492..451M}. The cause of this feature
  is that the projected inner rim emission area is null when the disc
  is seen face-on. In contrast, the hot-dust emitting area in the
  soft-edge models with rounded inner rim show much less dependency on
  the viewing angle.

At 45 degree inclination the flux at 3~$\mu$m
matches the fluxes from all the other disc-structure models. The flux
comes mostly from the edge facing the star, which cannot be seen at
null inclination. The SED between 8 and 100~$\mu$m is extremely
sensitive to the value of the flaring index $p$ because the more flare
a disc is, the more stellar photons are intercepted by the outer disc,
which predominately emits in the far infrared. The short wavelength
flux (3--8 $\mu$m) is mostly sensitive to the height of the rim, which
is constant, and changes only a little when we vary $p$.

SEDs alone cannot be used to differentiate between the inner disc
structures. On the other hand, images at short wavelengths (see
Fig.~\ref{fig_image_1.98_rim}) show differences between the models.
The emission at 1.98~$\mu$m of the fiducial model is concentrated in
the rounded inner rim with contribution from the second rim. The
emission of the fixed-disc model is spread over the first
AU. Discriminating disc models by fitting simultaneously the SED and
images has already been used successfully to study the disc around the
T~Tauri star IM~Lupi \citep{Pinte2008A&A...489..633P}.

\begin{figure*}  
\centering
  {\includegraphics[angle=0,width=18cm]{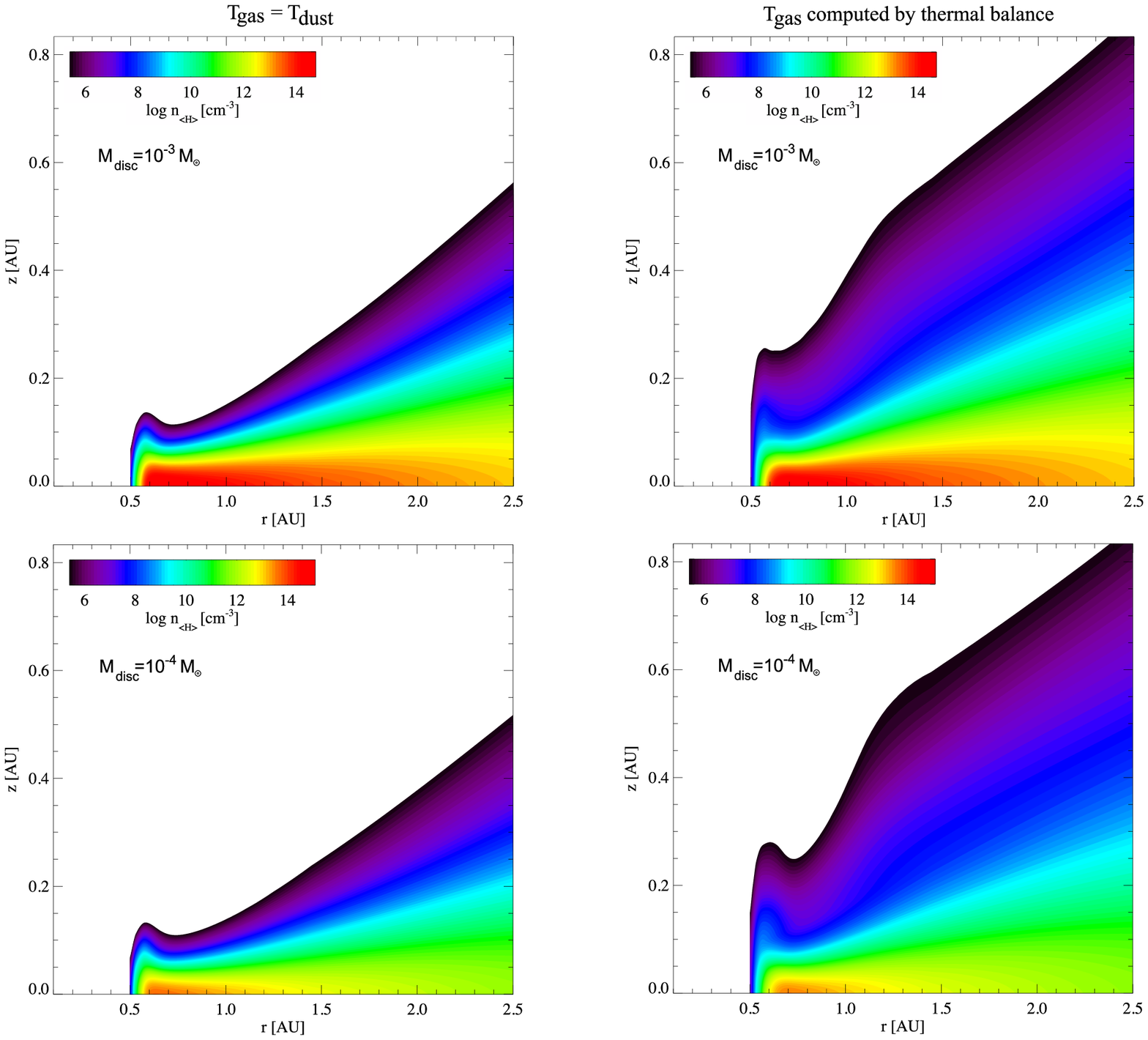}}
  \caption{Inner rim density structure for the $M_{\mathrm{disc}}$=10$^{-3}$ M$_\odot$ (upper panels) and $M_{\mathrm{disc}}$=10$^{-4}$ M$_\odot$ (lower panels). The left panels are the models with $T_{\mathrm{gas}}=T_{\mathrm{dust}}$. The right panels are the models with $T_{\mathrm{gas}}$ computed by gas thermal balance.}
  \label{fig_disc_mass_inner_rim}
\end{figure*}

\subsection{Testing the effect of ``soft'' versus ``sharp'' edge}\label{sharp_soft_edge}

Current disc models assume that the disc density decreases
monotonically from the inner edge to the outer radius (``sharp-edge''
model). We study the effect of adopting our ``soft-edge'' model
compared to ``sharp-edge'' model on the disc structure and
SEDs. Figure~\ref{inner_rim_structures} shows the density structure
for the fiducial disc parameters assuming
$T_{\mathrm{gas}}=T_{\mathrm{dust}}$ on the left panel and of
$T_{\mathrm{gas}}$ computed by gas thermal on the middle panel. Those
two density structures contrast with the two upper panels in
Fig.~\ref{fig_surf_density_inner_rim}.  

First the rim is extremely narrow in the ``sharp-edge'' models.  On the
other hand, the main effect of the 'soft-edge' model is to create a
rounded rim surface whatever the surface density profile (see
Fig.~\ref{inner_rim_structures}). The density gradients at the rim are
clearly seen in the density structure plots. The projected rounded rim
appears as an ellipse, similar to other theoretical studies
\citep{Isella2005A&A...438..899I,Kama2009A&A...506.1199K}.

In the ``sharp-edge'' models, the strength of the 3~$\mu$m excess
emission depends on the inclination of the disc with respect to the
observer. This dependency stems from the thinness of the rim and was
already found in previous studies of thermal-coupled models
\citep{Meijer2008A&A...492..451M}. On the other hand, the 3~$\mu$m
excess depends much less on the inclination for the ``soft-edge''
models (see upper-left panel of Fig.~\ref{fig_SEDs}).

Secondly, there is no secondary ``bump'' in the structure at
$\sim$1.2~AU, even when $T_{\mathrm{gas}}$ is computed by thermal
balance. The presence of the second bump in the ``soft-edge''
decoupled model can explain the difference in fluxes between the
models in the 4 and 40 micron region in the lower-right panel of
Figure~\ref{fig_SEDs}. The optically-thick rim in the ``sharp-edge''
models prevents a large amount of photons to reach the disc surface at
radius 1--1.5~AU (the so-called ``shadow''), therefore suppressing the
rise of the second bump. The bump exists because the gas stays at
temperatures $>$~1000~K up to 4~AU at the surface while the dust
temperatures are only $>$~300~K. The bump creates a large emitting
area for the warm dust, which translates into strong emission in the
4--40 $\mu$m region in the SEDs (See Fig.~\ref{fig_image_4.96_rim} and
upper-left panel of Fig.~\ref{fig_SEDs}). 

In the ``sharp-edge'' models, the mid-infrared flux is depressed
  compared to the ``soft-edge'' models because of the lower disc
  height behind the rim.

\begin{figure*}
\centering  
  {\includegraphics[angle=0,width=18cm]{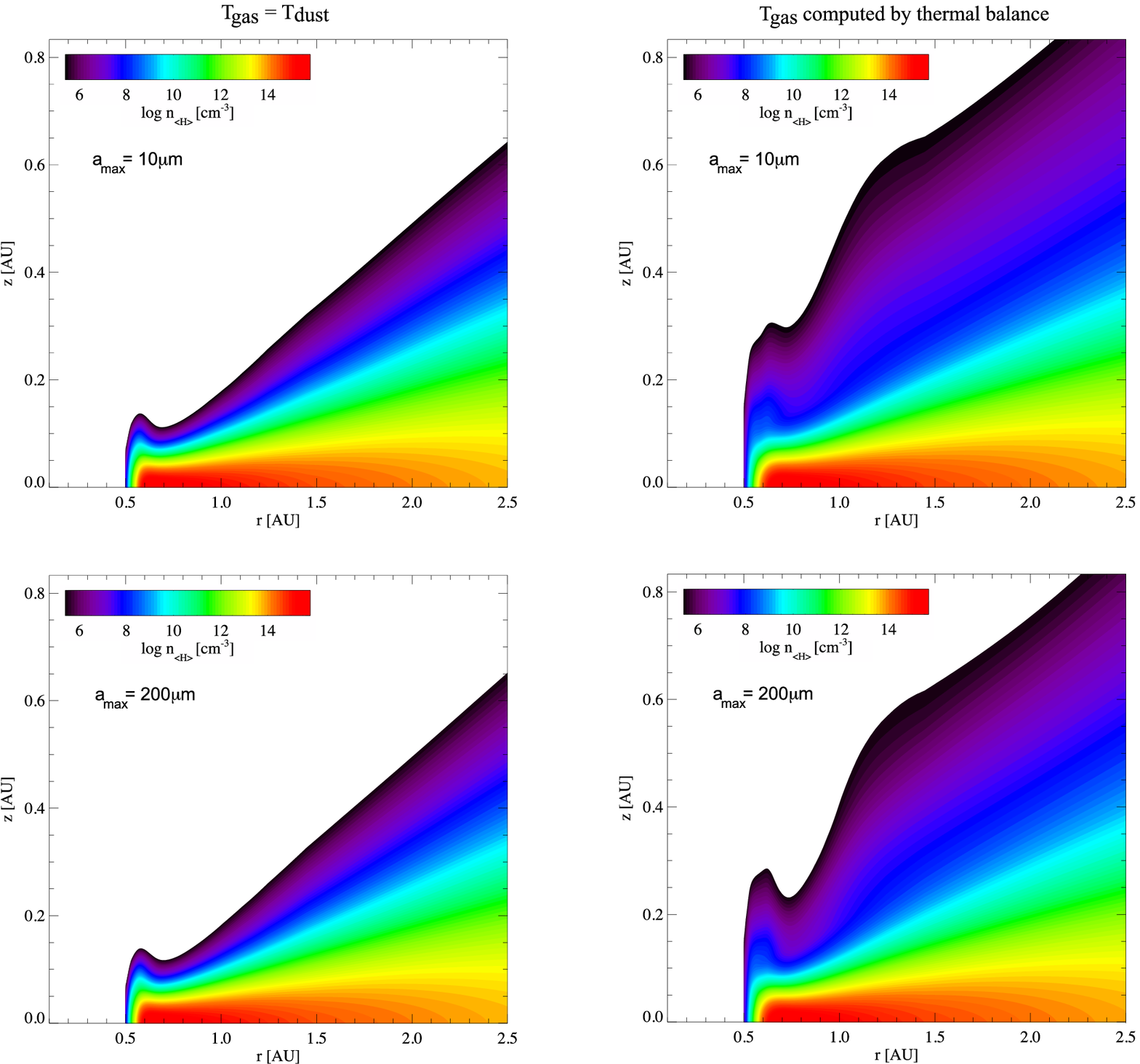}}
  \caption{Inner rim density structure for the
    $a_{\mathrm{max}}=$~10~$\mu$m and
    $a_{\mathrm{max}}=$~200~$\mu$m. The left panels are the models
    with $T_{\mathrm{gas}}=T_{\mathrm{dust}}$. The right panels are
    the models with $T_{\mathrm{gas}}$ computed by gas thermal
    balance.} 
  \label{fig_grain_size_structure}
\end{figure*}
\subsection{Effects of PAH abundance}\label{pah_effects}

\begin{figure*}   
\begin{minipage}[b]{0.48\linewidth}
\centering
{\includegraphics[scale=0.48]{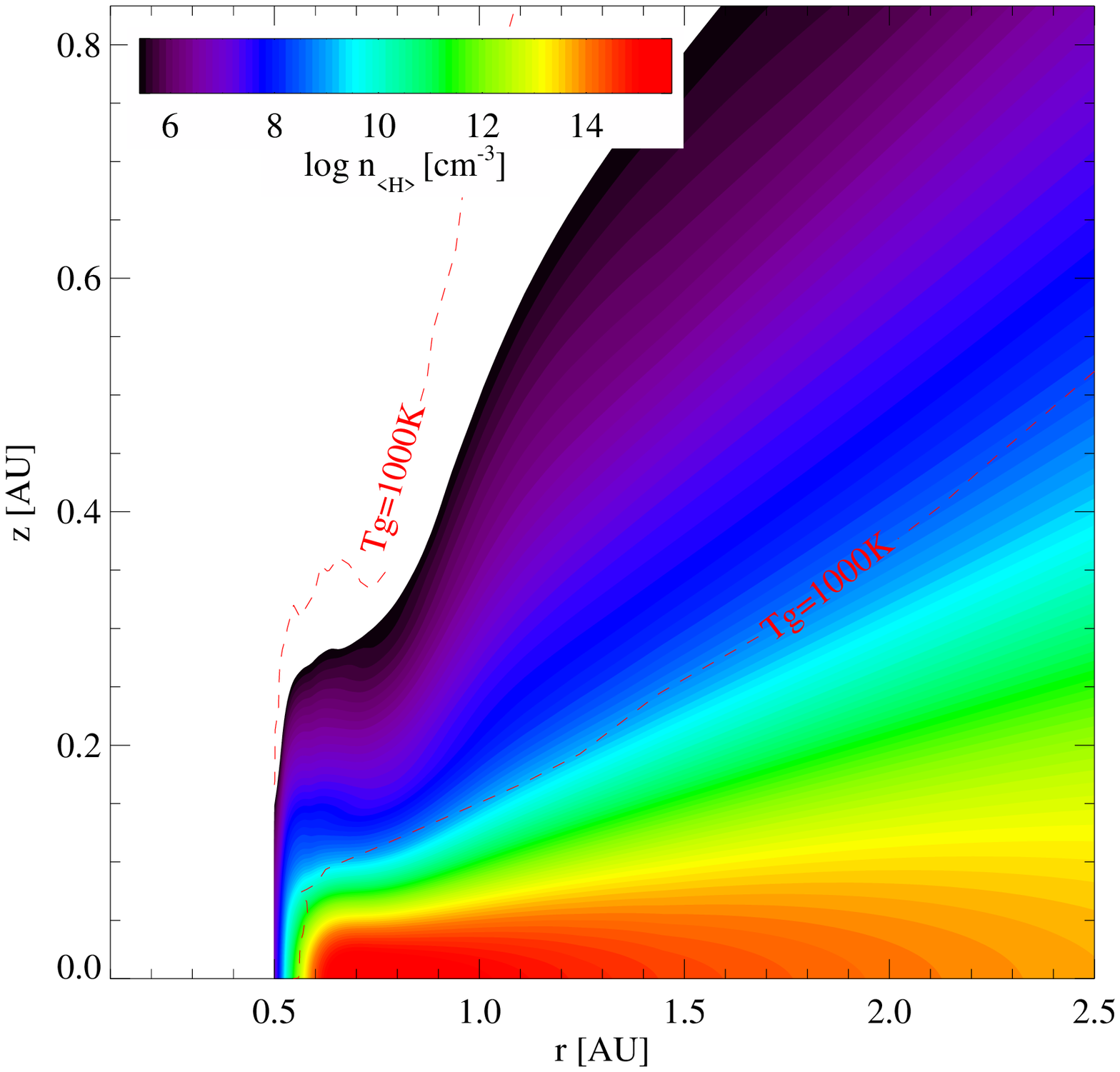}}
\end{minipage}
\begin{minipage}[b]{0.48\linewidth}
\centering
{\includegraphics[scale=0.48]{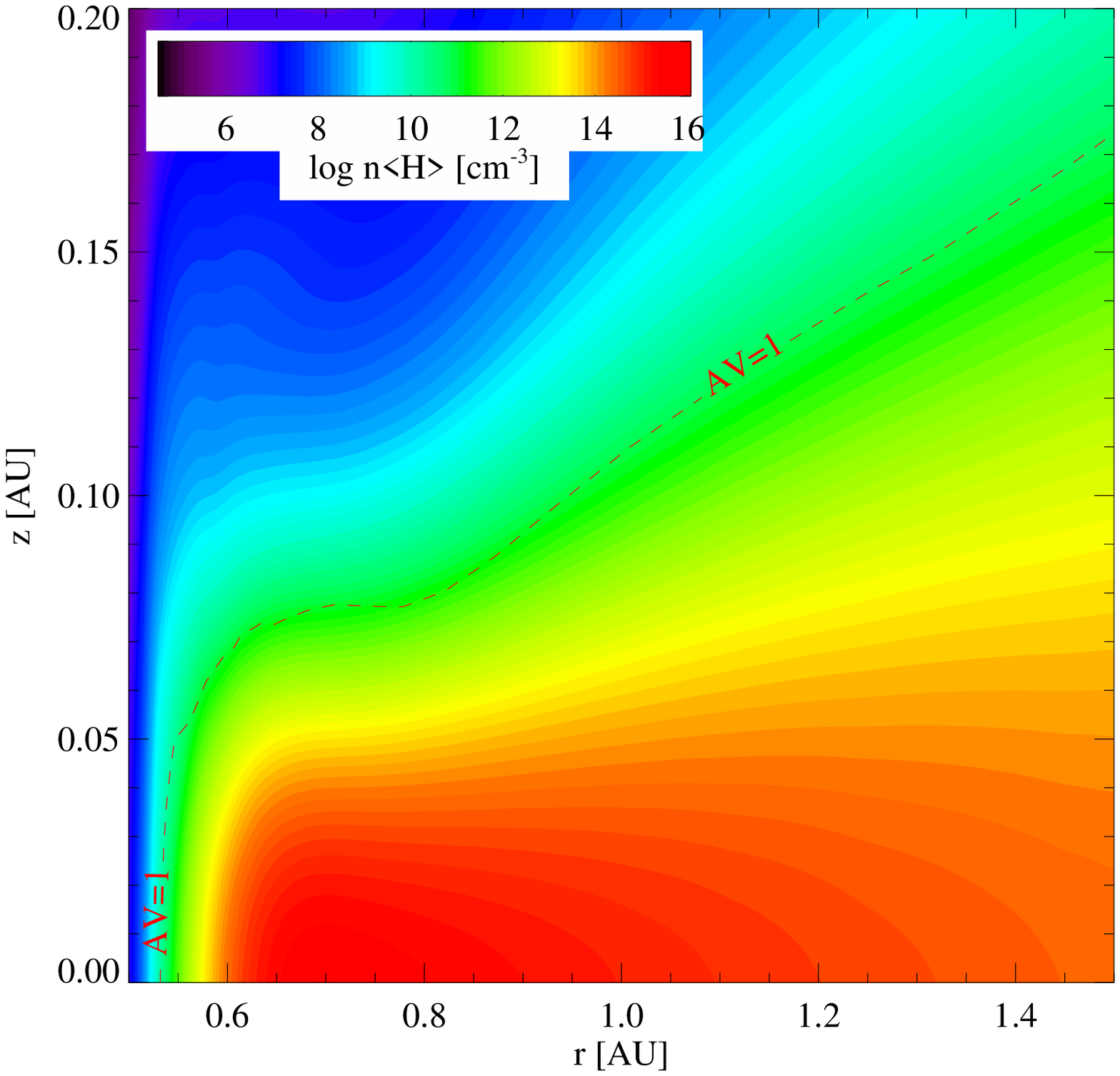}}
\end{minipage}
\begin{minipage}[b]{0.48\linewidth}
\centering
{\includegraphics[scale=0.48]{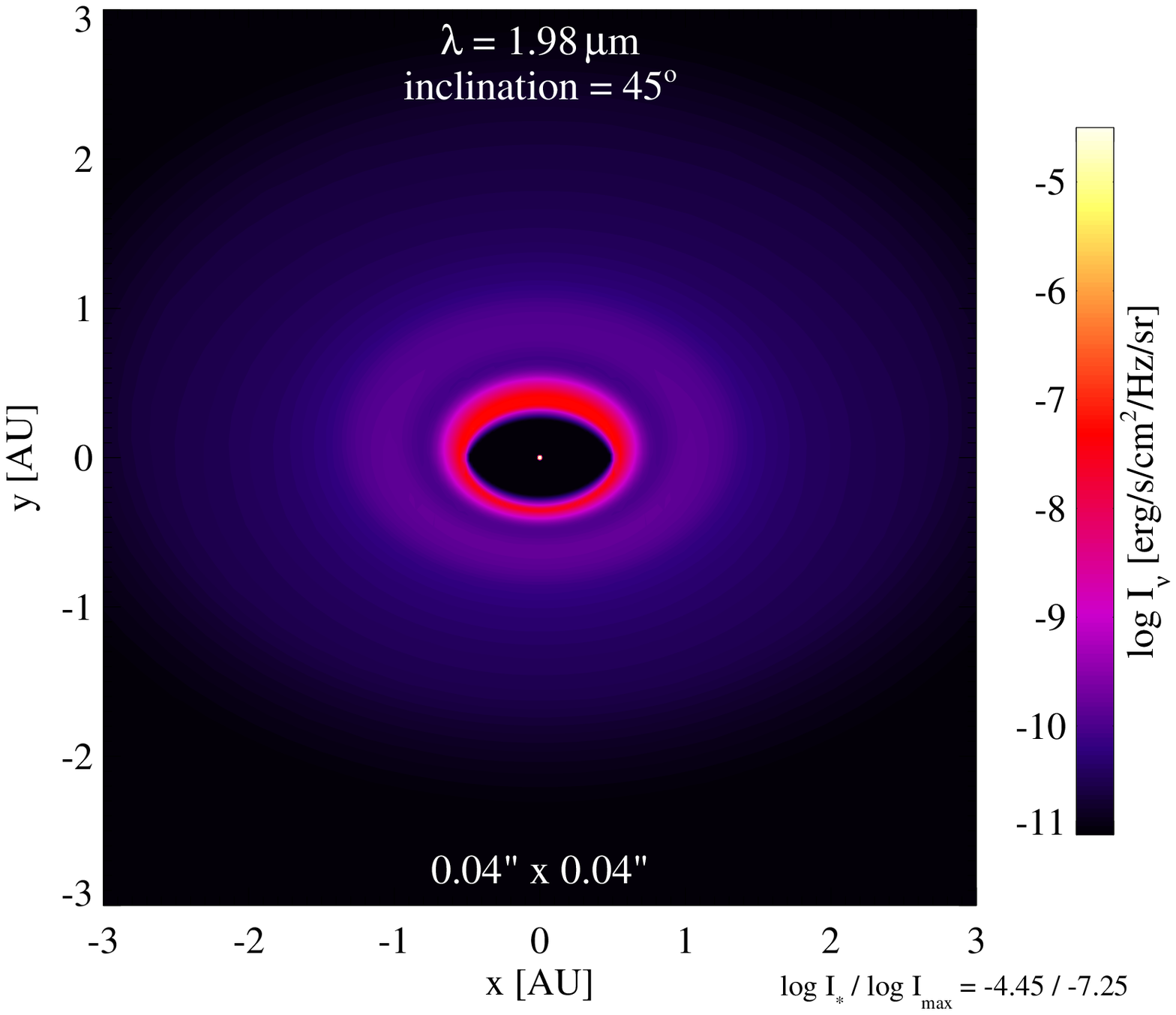}}
\end{minipage}
\begin{minipage}[b]{0.48\linewidth}
\centering
{\includegraphics[scale=0.48]{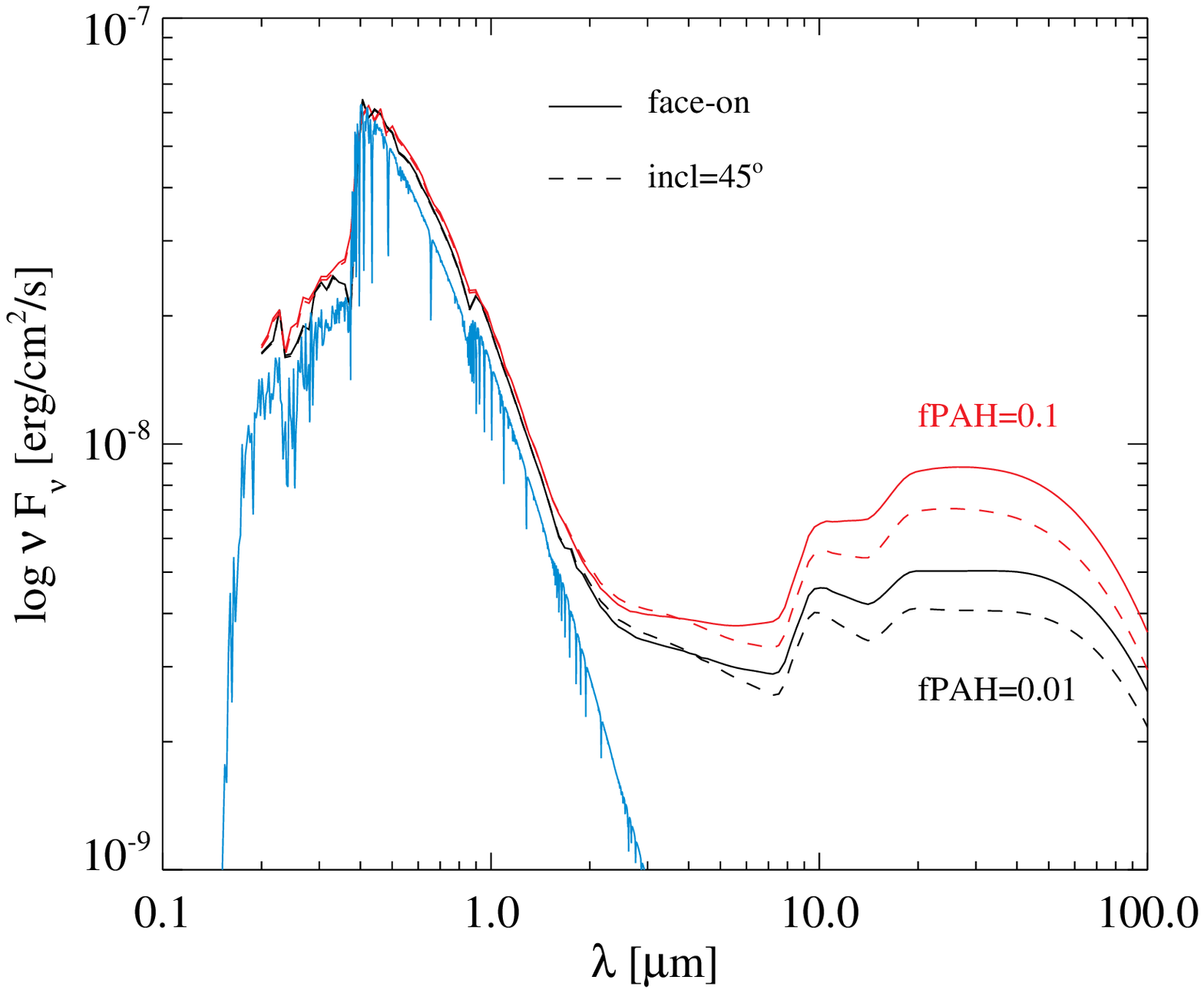}}
\end{minipage}
\caption{\label{fig_fPAH} Inner disc density structure (upper-left panel), photospheric height (upper-right panel), near-infrared image (lower-left panel), and comparison between the SED with  $f_{\mathrm{PAH}}$=0.01 and $f_{\mathrm{PAH}}$=0.1 for the fiducial model (lower-right panel). The red-dashed line in the upper-left panel encloses the region where the gas temperature is higher than 1000~K.}
\end{figure*}

The PAHs are the main heating of the agents of the gas
  \citep{Kamp2010A&A...510A..18K}. A disc with $f_{\mathrm{PAH}}$=0.1 is warmer than
a disc $f_{\mathrm{PAH}}$=0.01. A warmer disc is more extended vertically and intercepts
more radiation from the star (Fig.~\ref{fig_fPAH} upper-left compared to
Fig.~\ref{fig_surf_density_inner_rim} upper-right panel) although the 
shape of the photospheric height does not vary significantly
(Fig.~\ref{fig_fPAH} upper-right panel).
The overall emission in the continuum and in the gas lines is higher
(Fig.~\ref{fig_fPAH} lower-right panel). We also show an image
at 1.98~$\mu$m in the lower-left panel of Fig.~\ref{fig_fPAH}.

\subsection{Effects of varying the surface density profile}\label{surf_density_profile}

We show the results of varying the surface density power-law index
from $\epsilon=$~1.0 to $\epsilon=$~2.0. All other parameters are kept
constant. The density structure for the thermal-coupled
($T_{\mathrm{dust}}=T_{\mathrm{gas}}$) and thermal-decoupled
($T_{\mathrm{gas}}$ from thermal balance) models are displayed in
Fig.~\ref{fig_surf_density_inner_rim} for the inner disc on the right
and left respectively. The Spectral Energy Distribution for the models
are shown in the upper-left panel of Fig.~\ref{fig_SEDs}. 

The structure of thermal-decoupled disc models differ substantially
from their thermal-coupled counterparts.  The gas is heated by gaseous
and dust grain photoprocesses that convert ultraviolet photons into
fast-moving electrons, which in turn share the energy to the gas
(mostly atomic and molecular hydrogen). In the upper disc layers,
  the density is too low for efficient gas-grain thermal accommodation 
  and the gas and dust are thermally decoupled. The
  gas mostly cools by line emissions, which become quickly optically
  thick, while dust grains cool by optically thinner continuum
  emission. As a result, the gas remains at higher temperature than
  the dust grains.

Below the disc atmosphere, the ultraviolet flux is
attenuated and thermal accommodation between gas and dust grows. The
gas there is mostly molecular and frozen onto grain surfaces. Finally,
in the highly extinct midplane thermal accommodation dominates and
drives towards equal gas and dust temperatures. The dust temperature
structure is not affected by the change in density structure as there
are still two layers: a warm upper layer and a vertical isothermal
interior.

\begin{figure*}
\centering
  {\includegraphics[angle=0,width=18cm]{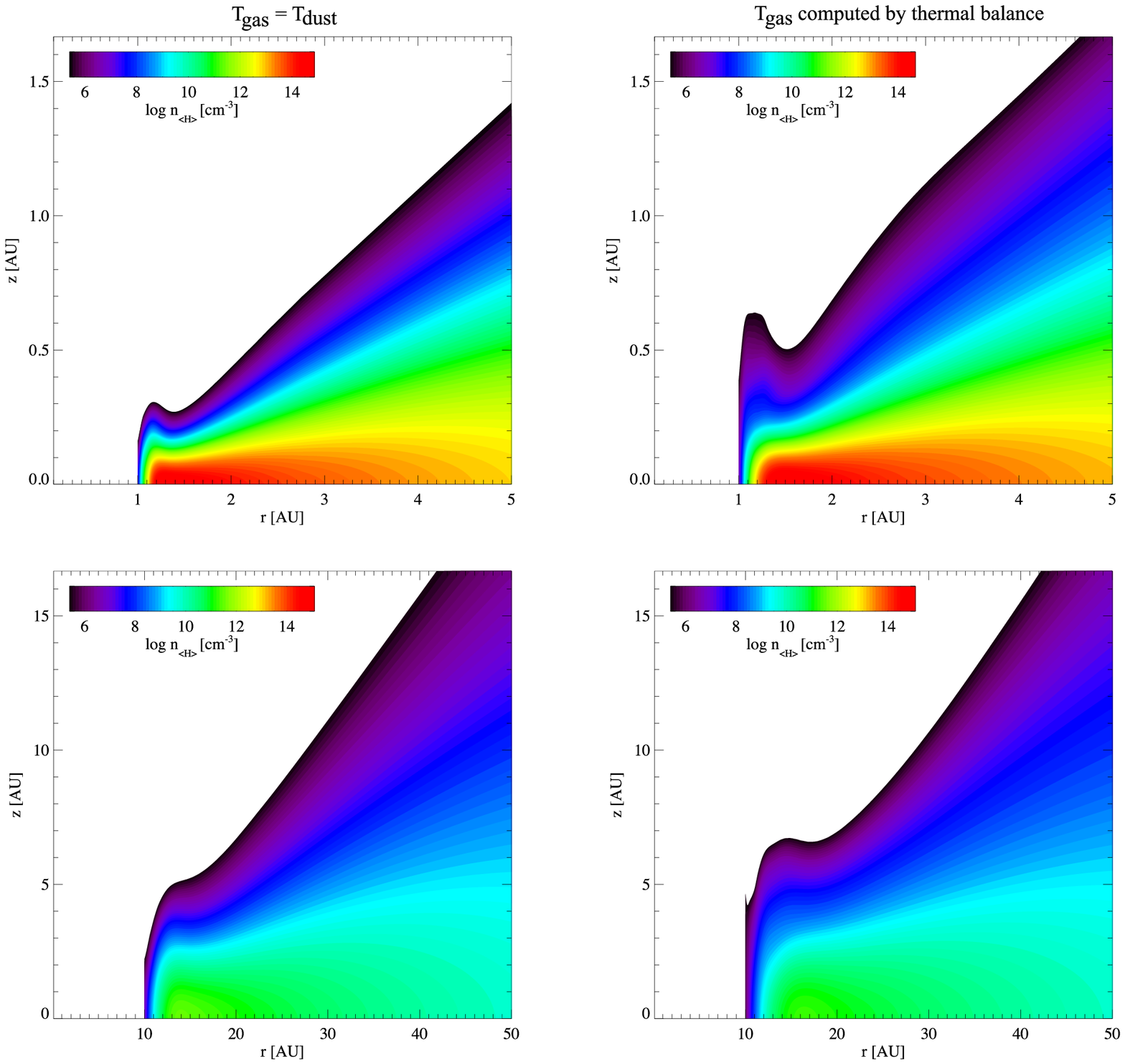}}
  \caption{Inner rim density structure for the $R_{\mathrm{in}}$=1
    (upper panels) and 10~AU (lower panels). The left panels are the
    models with $T_{\mathrm{gas}}=T_{\mathrm{dust}}$. The right panels
    are the models with $T_{\mathrm{gas}}$ computed by gas thermal
    balance.}
  \label{fig_inner_radius_rim}
\end{figure*}


The inner rim is much higher when the gas temperature is determined by
detailed heating and cooling balance.  The SED reflects relatively
well the amount of hot grains and thermal-decoupled models show a
larger amount because of higher rims.

In the thermal-coupled model, the rim is much lower and does not block
photons to reach the area behind the
rim. \citet{Acke2009A&A...502L..17A} studied the SED from Herbig~Ae
discs using a Monte-Carlo code coupled with hydrostatic equilibrium
assuming equal temperatures. They found good matches to observational
data when they artificially rise the height of the rim, consistent
with our results when the gas thermal balance is computed.

The second bump at 1--2~AU that appears in the thermal-decoupled
models are not present in the $T_{\mathrm{gas}}=T_{\mathrm{dust}}$
models (the left panels in Fig.~\ref{fig_surf_density_inner_rim}). As
already mentioned in the discussion of ``sharp-edge'' versus
``soft-edge'' models, the second bump is caused by the gas that is
hotter than the dust at the disc surfaces. A rounded optically thinner
rim helps more scattered light to reach the disc surfaces compared to
a narrow optically thicc rim.

The existence of a puffed-up rim is a necessary but not sufficient
condition for strong 3 $\mu$m excess. The surface area of the rim has
to be large, i.e. the rim has to be high enough. In the seminal paper
on disc inner rim by \citet{Dullemond2001ApJ...560..957D}, the height
of the rim to the rim distance from the star ratios were found to be
of the order of 0.1--0.25. We find that the rim in the case of
  $\epsilon=2.0$ is 0.25 AU high at 0.5~AU. Here we have defined the disc
  height as the location of the disc where gas density is 10$^{6}$
  cm$^{-3}$. A higher rim is needed in our modelling to produce
strong emission at 3$\mu$m because the
\citet{Dullemond2001ApJ...560..957D} model assumed a single dust
temperature at 1500~K for the rim while in our models only the dust in
the rim at the midplane reaches 1500~K. In the upper-left panel of
Fig.~\ref{fig_fiducial}, we can see a vertical decreasing dust
temperature gradient from the midplane (1500K) to the surface
($<$1000~K) at the rim radius.
  
The mid-infrared continuum emission at a specific wavelength reflects
the amount of dust grains at a temperature roughly given by Wien's
law. As found in previous studies, discs flare in the outer disc in
the thermal-coupled models. The rim deprives the disc atmosphere
behind the rim of UV photons. This shadowing effect has been invoked
to explain SED of disc with low 30--100 $\mu$m emission flux
\citep{Dullemond2004A&A...417..159D,Meijer2009A&A...496..741M,verhoeff2010A&A...516A..48V}.
This shadowing effect is less pronounced if dust scattering is taken
into account. Dust scattering ensures that some photons reach the
``shadowed'' area behind the rim.

The main effect of warmer upper layers in the decoupled models is
puffed-up and strongly flaring upper layers from the inner rim to the
outer radius. Although the inner rim is twice higher than in the
thermal-coupled case, the extra dust opacity in the rim is not
sufficient to attenuate the UV flux responsible for the gas heating
behind the rim. Denser gas are found higher in the disc and more
grains are raised to higher temperature, resulting in slightly
stronger 10--50 $\mu$m continuum flux. In general our hydrostatic
disc models flare strongly.

\subsection{Effects of varying the disc mass}\label{disc_mass}

In this series of models, the surface density profile is kept at 2.0
and we vary the disc mass ($M_{\mathrm{disc}}$=10$^{-4}$, 10$^{-3}$,
and 10$^{-2}$ M$_\odot$). The structure and SEDs for this series are
shown in Fig.~\ref{fig_disc_mass_inner_rim} and \ref{fig_SEDs}
respectively. The near-IR flux is weakly affected by the disc mass.
On the other hand, the flux in the 10--50 $\mu$m region is sensitive
to the total dust mass and thus decreases with decreasing total disc
mass (the gas-to-dust mass ratio is kept constant at 100), consistent
with the findings of \citet{Acke2009A&A...502L..17A}. From the SEDs in
Fig.~\ref{fig_SEDs}, the disc mass is the most important parameter
controlling the near- to far-IR flux ratio. As the opacity decreases
with wavelength (see Fig.~\ref{fig_opacity}), the flux depends less and
less on the disc geometry and more and more on the total dust mass,
which scales with the gas mass if a constant gas-to-dust mass ratio is
assumed.

\subsection{Effects of varying the maximum grain size}\label{big_grains}

The maximum grain size has a relatively weak influence on the inner
disc structure (Fig.~\ref{fig_grain_size_structure}). Therefore the
flux at 3--5~$\mu$m does not change significantly with increasing dust
grain upper size limit (see Fig.~\ref{fig_SEDs}). On the other hand,
the effects on the 10--70~$\mu$m shape of the SED are significant. The
bigger the grains are, the cooler they are and the emission is weaker
as testified by the higher opacity per unit mass for the small grains
\ref{fig_opacity}. In addition, big grains have less surface area per
volume and therefore the photoelectric heating efficiency
decreases. The disc is cooler and flares less, which results in less
intercepted stellar photons. The combined effects concur to decrease
the 10--70 flux when grains are big. Beyond 100~$\mu$m, the grain
emissivity depends strongly on the presence of larger grains.

At wavelength larger than their sizes, grains are efficient
emitters. Therefore the flux drop at wavelengths longer than 70~$\mu$m
is less pronounced for big grains. In summary, increasing the maximum
grain radius results in weaker emission at 30--70 $\mu$m but stronger
emission at longer wavelengths.

\subsection{Effects of varying the inner disc radius}\label{inner_radius}

Figure \ref{fig_SEDs} and \ref{fig_inner_radius_rim} show the SEDs and
inner disc structure for $R_{\mathrm{in}}$=~1 and 10~AU. A round rim
structure is present for all inner disc radii. Since the hottest
grains are cooler, the peak of the near-infrared bump has shifted to
$\sim$~4~$\mu$m in the $R_{\mathrm{in}}$=~1~AU model. The spectral
signature of the inner rim has disappeared for the
$R_{\mathrm{in}}$=~10~AU model. The presence of the puffed-up rim does
not depend on the actual value of the inner radius but only rims close
to the star have hot enough dust to emit strongly in the
near-infrared.

\section{Conclusions}\label{conclusion}

We modelled the structure, Spectral Energy Distribution, and infrared
images of protoplanetary discs around Herbig Ae star and analysed the
difference between the assumption of equal gas and dust temperatures
and when both temperatures are computed independently. We also
compared the hydrostatic disc model SEDs to the SEDs of a disc with
prescribed density structure.
  
The disc structure is governed by the gas pressure support, which in
turn depends on the gas molecular weight and temperature. The height
of the inner rim in models with calculated gas temperature exceeds
those where the gas and dust temperature are equal by a factor 2 to
3. Higher rims result in large emitting areas and thus slight larger
near-infrared excess. Our treatment of the inner disc density fall-off
(``soft-edge'') results in rounded inner rims consistent with
near-infrared interferometric studies. The discs also show a second
density bump that manifests itself as stronger emission between 3 and
30 $\mu$m. The flux beyond 30 $\mu$m is mostly sensitive to the disc
mass. The maximum grain radius affects weakly the SEDs. The effect of
gaps is to remove hot dust grains which emit predominately in the
near-IR. As a result, the flux in the near-IR is suppressed. The shape
of the SED from 3 to 100 $\mu$m cannot be used to discriminate between
the inner disc structure models. Together with the SED, images with
high spatial resolution may be used to differentiate between disc
models.

Our study stresses the importance of understanding the interplay
between the gas and the dust in protoplanetary discs. This interplay
shapes the disc structures, which in turn control the shape of the
SEDs.

\section{Acknowledgments}
WFT was supported by a Scottish Universities Physics Alliance (SUPA)
fellowship in Astrobiology at the University of Edinburgh. W.-F.\ Thi
acknowledges PNPS, CNES and ANR (contract ANR-07-BLAN-0221) for
financial support. We thank Ken Rice for discussions on inner disc
structures.

\bibliographystyle{mn2e}
\bibliography{rim_herbigAe}  
\bsp

\label{lastpage}

\end{document}